\documentclass[a4paper,useAMS,usenatbib]{mn2e}

\usepackage{graphicx}
\usepackage{epstopdf}
\usepackage{subfig}
\usepackage{natbib}
\usepackage{xcolor}

\usepackage{amssymb}

\title[Modeling Interacting Galaxy Pairs - Testing Identikit]{Modeling the Initial Conditions of Interacting Galaxy Pairs Using Identikit}
\author[S. Alireza Mortazavi, Jennifer M. Lotz, Joshua E. Barnes, Gregory F. Snyder]{S. Alireza Mortazavi$^{1}$\thanks{E-mail:
alireza@pha.jhu.edu}
, Jennifer M. Lotz$^{2}$
, Joshua E. Barnes$^{3,4}$
, Gregory F. Snyder$^{2}$ \\
$^{1}$Department of Physics and Astronomy, Johns Hopkins University, 3400 N. Charles St, Baltimore, MD 21218, USA\\
$^{2}$Space Telescope Science Institute, 3700 San Martin Dr., Baltimore, MD 21218, USA\\
$^{3}$Institute of Astronomy, University of Hawaii, 2680 Woodlawn Drive, Honolulu, HI 96822, USA\\
$^{4}$Yukawa Institute for Theoretical Physics, Kyoto University, Kitashirakawa Oiwakecho, Sakyo-ku, Kyoto 606-8502 Japan}
\begin{document}

\date{Accepted ???? ???? ??. Received ???? ???? ??; in original form 2014 December 17}

\pagerange{\pageref{firstpage}--\pageref{lastpage}} \pubyear{2015}

\maketitle

\label{firstpage}

\begin{abstract}
We develop and test an automated technique to model the dynamics of interacting galaxy pairs. We use Identikit (\citealt{Barnes:2009fh}; \citealt{Barnes:2011kb}) as a tool for modeling and matching the morphology and kinematics of the interacting pairs of equal-mass galaxies. In order to reduce the effect of subjective human judgement, we automate the selection of phase-space regions used to match simulations to data, and we explore how selection of these regions affects the random uncertainties of parameters in the best-fit model. In this work, we use an independent set of GADGET SPH simulations as input data to determine the systematic bias in the measured encounter parameters based on the known initial conditions of these simulations. We test both cold gas and young stellar components in the GADGET simulations to explore the effect of choosing HI vs. H$\alpha$ as the line of sight velocity tracer. We find that we can group the results into tests with good, fair, and poor convergence based on the distribution of parameters of models close to the best-fit model. For tests with good and fair convergence, we rule out large fractions of parameter space and recover merger stage, eccentricity, pericentric distance, viewing angle, and initial disc orientations within 3$\sigma$ of the correct value. All of tests on prograde-prograde systems have either good or fair convergence. The results of tests on edge-on discs are less biased than face-on tests. Retrograde and polar systems do not converge and may require constraints from regions other than the tidal tails and bridges.

\end{abstract}

\begin{keywords}
galaxies: kinematics and dynamics, galaxies: interactions
\end{keywords}

\section{Introduction}

The merger of galaxies and their associated dark matter halos is a fundamental process in galaxy evolution and cosmology. Galaxies and the dark matter halos they live in grow in time through mergers and accretion from the cosmic web. Over the past 8 billion years massive galaxies have transformed from star-forming disc galaxies into inactive bulge-dominated ones (e.g. \citealt{Bell:2004fo}; \citealt{Faber:2007bt}). Galaxy mergers may be an important process that drives the assembly of galaxies, rapid star formation at early times, and the formation of bulge-dominated galaxies (e.g. \citealt{Toomre:1972ji,Mihos:1996bo,Barnes:1996bn}). 

Direct measurements of the initial orbital conditions of colliding galaxies are useful constraints for cosmology and galaxy evolution. Cosmological dark matter simulations predict hierarchical gravitational growth of structure with time. Numerical simulations predict the distribution of orbital parameters of dark matter halo mergers (e.g. \citealt{Benson:2005hi}; and \citealt{Khochfar:2006de}). If galaxies follow the dark matter halos, direct measurement of orbital parameters in galaxy mergers will put constraints on these simulations. In addition, idealized and cosmological zoom-in galaxy merger simulations have shown that merger induced star-formation history may depend on the initial orbital parameters of the interacting pair. (e.g. \citealt{Cox:2008jj}; \citealt{Snyder:2011fs}). 

Moreover, integral field spectroscopy of nearby elliptical galaxies have shown that early-type galaxies in local universe can be categorized into fast and slow rotators (\citealt{Cappellari:2011ej}; \citealt{Krajnovic:2011jj}; \citealt{Emsellem:2011br}). Recent results have shown that initial orbital parameters of major galaxy mergers can affect the rotational properties of their bulge-dominated remnants (e.g. see \citealt{Hopkins:2009jj}, \citealt{Bois:2011kc}, \citealt{Naab:2014ht}). Measuring these orbital parameters in early-stage mergers will make it possible to predict the kinematic properties of the merger remnant. 

Constraining the initial conditions of a pair of interacting galaxies can be accomplished by finding a simulation reproducing both the morphology and kinematics of the data. (e.g. \citealt{Toomre:1972ji}; \citealt{White:1978ub}; \citealt{Barnes:1988bz}). While some efforts have been made to model interacting galaxies by matching only the morphology (\citealt{Toomre:1972ji}; \citealt{Shamir:2013uf}), line of sight velocity data is required to find unique dynamical models for many interacting systems (\citealt{Barnes:2009fh}). For example, the best-fit dynamical model for NGC 7252 (\citealt{Borne:1991fo}) changed significantly when high quality HI kinematics data became available (\citealt{Hibbard:1994kh}; \citealt{Hibbard:1995iz}). 

\cite{Barnes:2009fh} provides a review of the dynamical modeling of the interacting disc galaxies which have made use of kinematic information. These attempts have used different amounts of kinematic data; some have tried to match 2D kinematics obtained from HI or H$\alpha$ maps (e.g. \citealt{Hibbard:1995iz}; \citealt{Duc:2000is}; \citealt{Struck:2005eb}), while others have only used 1D kinematics from long-slit spectroscopy (e.g. \citealt{Mihos:1993kb}; \citealt{Diaz:2000fz}; \citealt{Scharwchter:2004iy}). Most of these attempts rely on human expert judgment about the model that best matches the data. There has been some attempts to automate the matching process using genetic algorithms (e.g. \citealt{Theis:2001dc}; \citealt{Wahde:2001fb}; \citealt{2003Ap&SS.284..495T}). These algorithms have not yet matured enough to replace visual matching. 

More kinematics information for modeling the dynamics of galaxy mergers will become available in near future with optical IFU and radio surveys providing large amounts of 2D line of sight velocity data for nearby galaxies. CALIFA (\citealt{Sanchez:2012ku}), SAMI (\citealt{Croom:2012fo}), and MaNGA (\citealt{2015ApJ...798....7B}) are ongoing surveys of optical 2D spectroscopy of nearby galaxies including large numbers of interacting galaxy pairs. In addition, Australian SKA Pathfinder (\citealt{Johnston:2008gq}) and MeerKAT (\citealt{Booth:2009wx}) will perform high resolution HI surveys of nearby universe. Thus, we need to develop robust tools to classify interacting galaxies based on this data.

Identikit is a tool for modeling major galaxy mergers (\citealt{Barnes:2009fh}; \citealt{Barnes:2011kb}). It combines self-consistent and test particle techniques in order to utilize fast exploration of the parameter space of a disc-disc encounter. With Identikit 1 (\citealt{Barnes:2009fh}), the user can interactively change parameters like viewing direction and the orientation of the two discs until the best visual match between model and data is found. This interactive interface has been used for dynamical modeling of some major galaxy mergers (\citealt{Privon:2013fs}). The visual match, though, is subjective and depends on user's judgment about the most similar model. It requires a great deal of human-expert time spent on exploring the parameter space and looking for the best match. More importantly, the uncertainty in the initial conditions measured with Identikit 1 is not determined. Identikit 2, however, defines a quantity called ``score" that provides an informal measure for the quality of the match. As a result, we can automatically search the parameter space and find the model with the maximum score, i.e. the best-fit model. 

In this paper, we developed an automated routine based on Identikit 2. Our goal is to test the random and systematic uncertainties of modeling a major galaxy merger system using our method. In order to measure the systematics of Identikit modeling, we used GADGET SPH simulations (\citealt{Cox:2006kb}; \citealt{Cox:2008jj}) as input data. Because the initial conditions of the GADGET simulations are known, we can measure the biases in the parameters of the best-fit model. We test both cold gas and young stars in the the GADGET simulations to compare the effect of using HI vs. H$\alpha$ as the kinematics tracer. Through a statistical approach, we also measure the random uncertainty of the best-fit model. In \S 2 and \S 3 we describe the methodology and the hydrodynamical simulations we use in this work. In  \S 4 and \S 5 we present the results and discussion respectively.

\begin{table*}
\centering
\begin{tabular}{lll}
	\hline
	Parameter Class		&	Parameter							& Range Tested											\\ \hline
	orbital parameters 		&	eccentricity(1)  						& [0.60-1.10] 												\\
						&	pericentric distance(1)				& [0.03125-1.0000]$\times R_{vir}$ 								\\ 
						&	mass ratio	(1)						& 1 														\\ \hline
	observer dependent 		&	time since pericenter(1)				& from first pass to second pass									\\
	parameters			&	viewing angle(2)					& found through maximizing score in Identikit 2						\\
						&	position(2)							& set by locking the centers									\\
						&	length scaling $\mathcal{L}$(1)			& set by viewing angle and locking the centers						\\
						&	velocity offset(1)					& 0														\\
						&	velocity scaling $\mathcal{V}$(1)		& [-0.500-+0.500]$^*$  											\\ \hline
	initial orientation		&	(4)								& found through maximizing score in Identikit 2						\\
	of discs				&									&														\\ \hline
\end{tabular}
\caption{List of Identikit parameters for a system of interacting galaxies. 
\protect\cite{Barnes:2011kb} categorized these parameter into 3 classes. The number
 of degrees of freedom in each parameter is shown in the parentheses. As
  discussed in the text, we test 4 of these parameters, eccentricity, pericentric
   distance, time since pericenter, and velocity scaling. For every Identikit model
    and for each value of velocity scaling Identikit 2 finds the best viewing angle
     and the best initial disc orientations by maximizing score. The rest of the
      parameters can by estimated from direct observations, independent of modeling.  
      $^*$ The velocity scaling $\mathcal{V}$ relates the dimensionless velocity in Identikit to the 
      physical velocity. If $V_{range}$ is the velocity range in the Identikit velocity
      panels (top right and bottom left panels in Figure \ref{fig:Identikit}), the physical scale of
      unit velocity in Identikit, $V_{phy}$, is $V_{range}/(4\times 10^{\mathcal{V}})$. }
\label{tab:Identikit}
\end{table*}

\section[]{Method}
\label{sec:method}

\subsection{Identikit}
\label{sec:identikit}

Identikit (\citealt{Barnes:2009fh} and \citealt{Barnes:2011kb}) matches simulated disc-disc galaxy encounters to the observed morphology and kinematics of disc-disc galaxy mergers. Assuming a particular mass model for the isolated galaxies (see \S \ref{sec:model_details}), there are 3 groups of parameters describing the encounter of two disc galaxies. First are the orbit parameters, including eccentricity, pericentric distance, and mass ratio. Second are the parameters that describe the initial angular momenta of discs. The last group contains the parameters that depend on the observer's time and location.The third group includes the viewing direction and the time of the merger as well as the parameters transforming the model to the physical scales of the real system (i.e. length/velocity scaling, and length/velocity offset). All of these make up dynamical parameters involved in a disc-disc galaxy interaction. We list these parameters in Table \ref{tab:Identikit}.

The isolated galaxies consist of massive self-consistent and massless test particles with no dissipative (gaseous) component. Each galaxy contains 81,920 massive particles and 262,144 massless test particles.The massive particles are distributed in a spherically symmetric fashion to represent the potential of a massive dark matter halo, a disc, and a bulge.  The massless test particles, on the other hand, are distributed in discs with circular orbits. The motion of test particles is governed by the gravitational potential produced by massive particles. However, because they are massless, one may simulate test particles for multiple discs with different orientations simultaneously, without having them affect each other. As a result, one can calculate the trajectories of all possible disc particles in a single simulation run.  In one run, the user obtains the morphology and kinematics of merger systems with discs of all different initial orientations. When comparing simulations to data, at each time step of the simulation, the user can turn on a particular disc and turn off the rest. Thus, Identikit quickly explores the parameter space in search for the best model.

Identikit can model the large scale morphology and kinematics of the tidal tails with a unique set of initial conditions. Early simulations of disc-disc galaxy mergers show that the shape and size of the tidal tails are sensitive to the initial conditions of the encounter (\citealt{Toomre:1972ji}; \citealt{Hibbard:1995iz}). These features move ballistically after the first passage and carry a memory of the initial conditions, and it's been shown that there is little difference in the shape of tidal features in test-particle vs. self-consistent simulations of interacting galaxy pairs (\citealt{1999ApJ...526..607D}). On the other hand, self-gravitating features (e.g. the spiral arms, stellar clusters) should not be matched when using a test-particle model like Identikit.

\begin{figure*}
\includegraphics[width=0.7\textwidth]{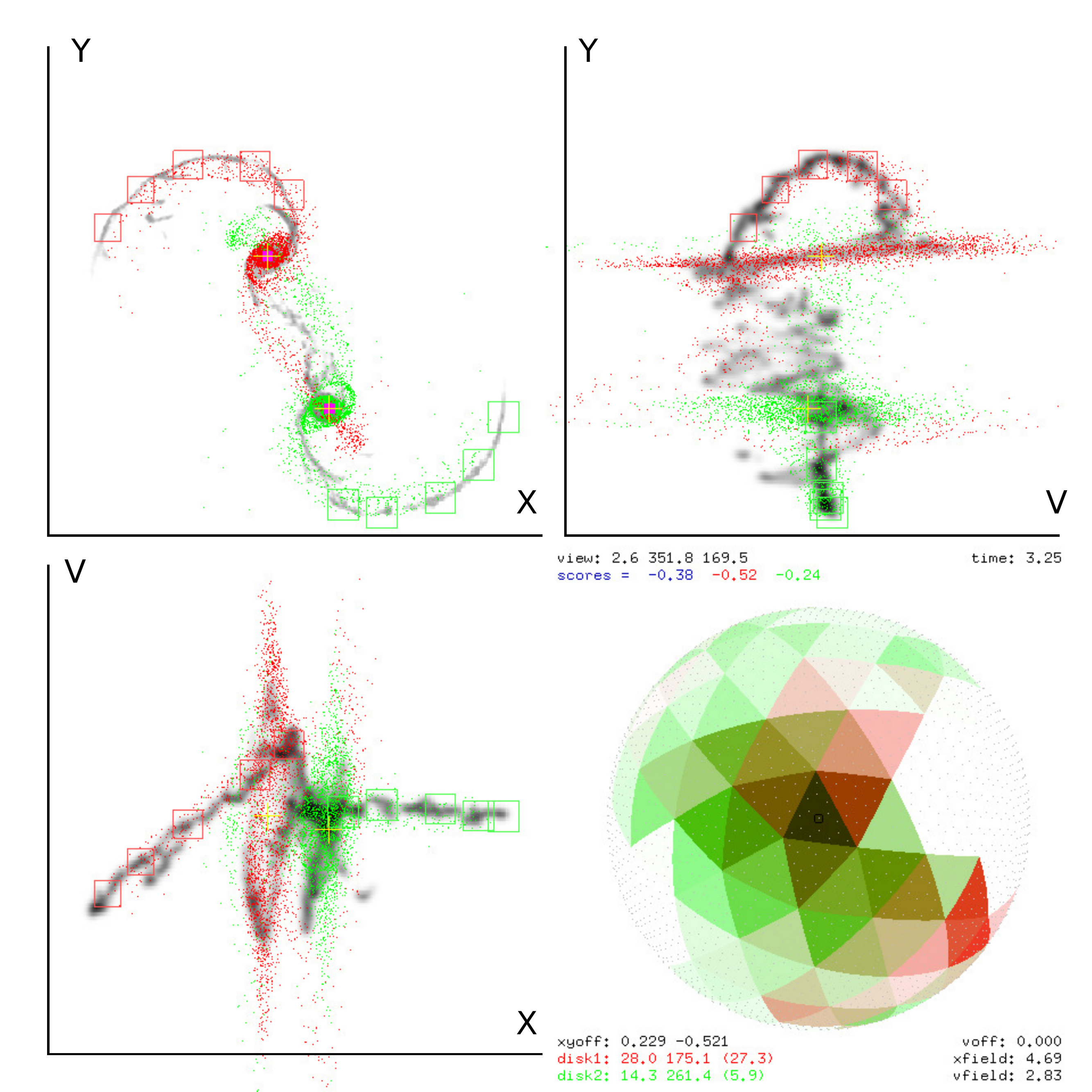}
\caption{ Identikit model fit to a GADGET simulated data. The top left panels show the 
morphology, the bottom left and top right panels show the LOS velocity-position.
The GADGET cold gas is shaded grey,  and the best-fit Identikit galaxy models 
are the red and green points. Identikit uses the phase-space regions selected 
on the tidal tails (red and green boxes) to calculate the quality of the fit for each
disc as functions of different viewing directions. Identikit calculates these functions at 320
viewing angles. The bottom right panel shows these functions as green 
and red shades on a spherical lattice. The best viewing direction is 
where the product of these functions are maximum.}
\label{fig:Identikit}
\end{figure*}

While Identikit 1 (\citealt{Barnes:2009fh}) identifies the best model interactively and qualitatively, Identikit 2 (\citealt{Barnes:2011kb}) introduces a score that quantifies the agreement between the model and the data. The score is calculated using user-input boxes which indicate regions of phase space and are extended in X, Y (morphology), and V (line of sight velocity) directions. As is shown in Figure \ref{fig:Identikit}, boxes are put on the tidal tails and bridges of the merger system. Identikit 2 calculates the scores based on the number of test particles populating these boxes. The score is assumed to be higher for the models that better reproduce the chosen phase-space boxes. Identikit 2 then scans viewing directions and initial disc orientations, calculating the score for each direction/orientation. The best orientation and viewing angles are the ones that give maximum score for a particular set of parameters.

The primary advantage of Identikit's test particle technique is rapid scanning of the multidimensional parameter space. Moreover, Identikit 2 quantifies the quality of the match, which is a unique feature. While visual matching is time consuming and subjective, providing a quantity (score) for the goodness of the fit is quantitative and requires less human expert time. However, massless test particles cannot reproduce self-gravitating features (e.g. spiral arms), and one cannot expect the dissipative features in gaseous components to be recovered. Tidal features are the key to recover a reasonable model. In some interacting galaxies such as retrograde ones, the tidal features are weak, and therefore they are more difficult to model. Besides, score does not have an absolute significance like $\chi^2$. It is only useful for comparing quality of fit with the same set of boxes and cannot be used independently to determine the likelihood of the fit.

We create a library of Identikit models consisting of different time steps of simulations of galaxy mergers with varying orbital parameters. Each member of this library is an individual frame with four fixed encounter parameters which we call external parameters (time, eccentricity, pericentric distance, and mass ratio). The user can study each member of the library in an interactive interface and explore other parameters of interest which include viewing direction, initial orientation of discs, length and velocity offsets, and length and velocity scalings. We call these internal parameters. Identikit 2 measures the score after the user locks the centers of model galaxies on the sky positions of the centers of the observed interacting galaxies. This constrains the position (length offset) and makes the viewing angle determine the length scaling $\mathcal{L}$ due to projection effect. So, locking the centers constrains two of the internal parameters. Identikit explores the viewing direction and the initial orientation of discs and finds the configuration that maximizes the score. So, for each member of the Identikit library at a particular velocity scaling and velocity offset, Identikit calculates a score.

In this work, merger systems consisting of two separate galaxies with distinct cores and strong tidal tails are examined.  Identikit 2 can only be used to model separate galaxies. Additionally, by selecting galaxies which have not yet coalesced, we can independently estimate the mass ratio based on the measured light ratio. In this paper, we only study equal mass galaxy mergers. Additionally, tidal features are strongest during the time range between the first and the second passage. So, we restrict the current study to this time range. It may be possible to estimate a prior value for the velocity offset  $\mathcal{V}$ by measuring the light-(mass-)weighted average (systemic) velocity of the merger system. Locking the centers constrains position and length scaling. As a result, ignoring the freedom in choosing the mass model in isolated galaxies, we have six more encounter parameters to explore: eccentricity, pericentric distance, time (between the first and second passage), viewing angle, orientation of discs, and velocity scaling. 

Of the remaining internal parameters, viewing angle and orientation of discs are determined when we maximize score for each member of the library; however, velocity scaling is a free parameter. We can find the best score for each member of Identikit library with a fixed velocity scaling. So, we calculate score for models with a grid of parameter values for eccentricity, pericentric distance, time since pericenter, and velocity scaling. Initial orientation of discs and viewing angle (and therefore length scaling  $\mathcal{L}$) are determined independently when we calculate the score for each of these models. Table \ref{tab:Identikit} shows the list of all parameters involved in Identikit and the range of the four parameters that are systematically explored in this work.

For an interacting system (i.e. a GADGET simulation in this work), one can make the map of scores of Identikit models with varying encounter parameters. In order to make such a map, we select a set of boxes on tides and bridges of the interacting system. Keeping the selected boxes fixed in place, we match all models in the Identikit library. For each member we change the velocity scaling as a free parameter and record how the score changes. Eventually, we obtain a score for all Identikit models with different eccentricity, pericentric distance, time, and velocity scaling. This makes an 4+1 dimensional scalar field. The model with the maximum score is the best-fit model. However, we still need to know how significant is the maximum score we found.

\begin{figure}
\includegraphics[width=0.45\textwidth]{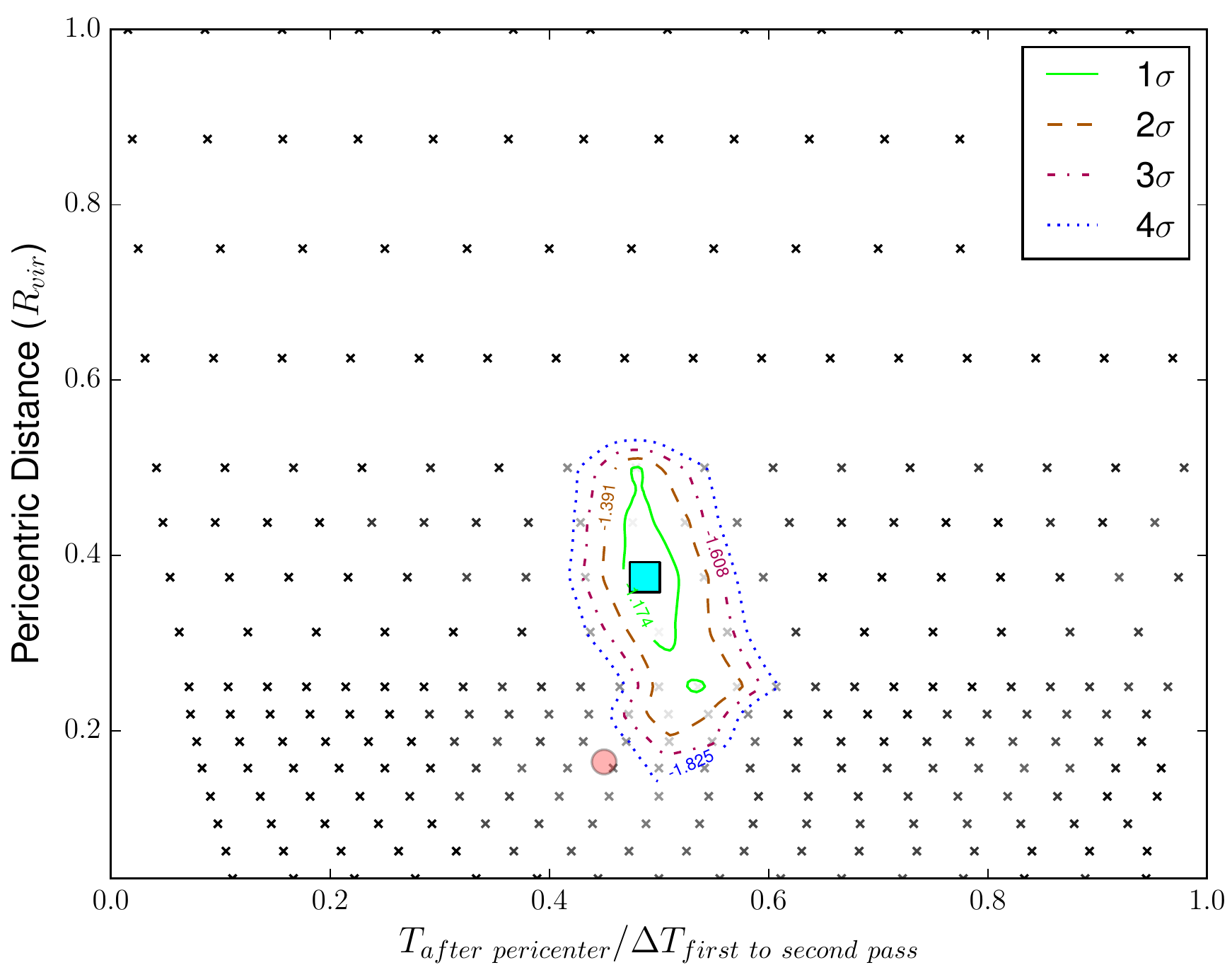}
\caption{The average score map of Identikit models 
matching Sbc201a (test 01). Each cross represents an 
Identikit model for which the score is calculated several
 times by moving the boxes (see \S \ref{sec:box_selection}).
 The average and standard deviation of the scores are calculated.
 The model with the highest average score is shown with
 the cyan box. The counters enclose the models with scores within
 1, 2, 3, and 4 times the standard deviation of 
 the best-fit model. The crosses are also color coded with 
 light gray showing higher scores and dark gray showing 
 lower scores. Notice that this figure shows a 2D
slice of the score map showing scores for models with varying
time and pericentric distance but fixed eccentricity (=1.0) 
and velocity scaling (-0.1). The red circle shows the known 
correct answer for the GADGET simulation.}
\label{fig:scoremap}
\end{figure}

We perform a statistical evaluation to measure the uncertainty of the score for each Identikit model. We select the same number of  boxes on the same tides and bridges but at slightly randomized positions. With the new set of boxes we calculate scores for all Identikit models again obtaining a new 4+1 dimensional score map. The new score map will have slightly different scores at each point in the parameter space. We repeat to make multiple score maps, each by moving the boxes around the tides and bridges of the interacting system. We calculate the average and the standard deviation of the scores at each point in parameter space. Figure \ref{fig:scoremap} shows a 2+1 dimensional slice of an average score map. Each cross represents an Identikit model with a particular set of parameters. The cyan box in Figure \ref{fig:scoremap} shows the best-fit model with the highest average score, and the red circle shows the input GADGET parameters. The contours show the models with scores that are within 1, 2, 3, and 4 standard deviation of the scores of the best-fit model. We calculate the uncertainty in the best-fit parameters from these contours.

\subsection{Box Selection}
\label{sec:box_selection}

\begin{figure*}
\centering
\subfloat[]{\label{fig:boxing1}\includegraphics[width=0.4\textwidth]{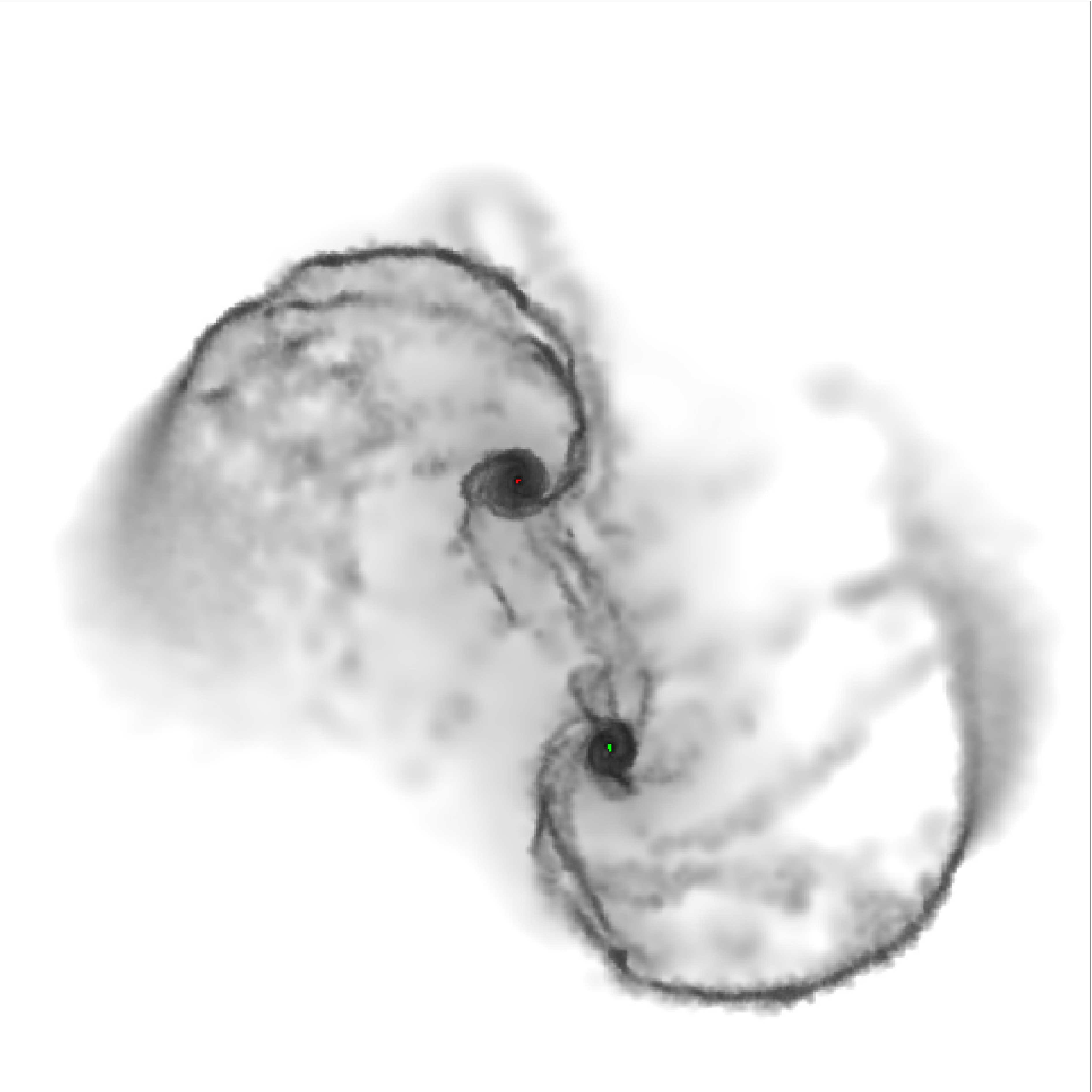}}  
\subfloat[]{\label{fig:boxing2}\includegraphics[width=0.4\textwidth]{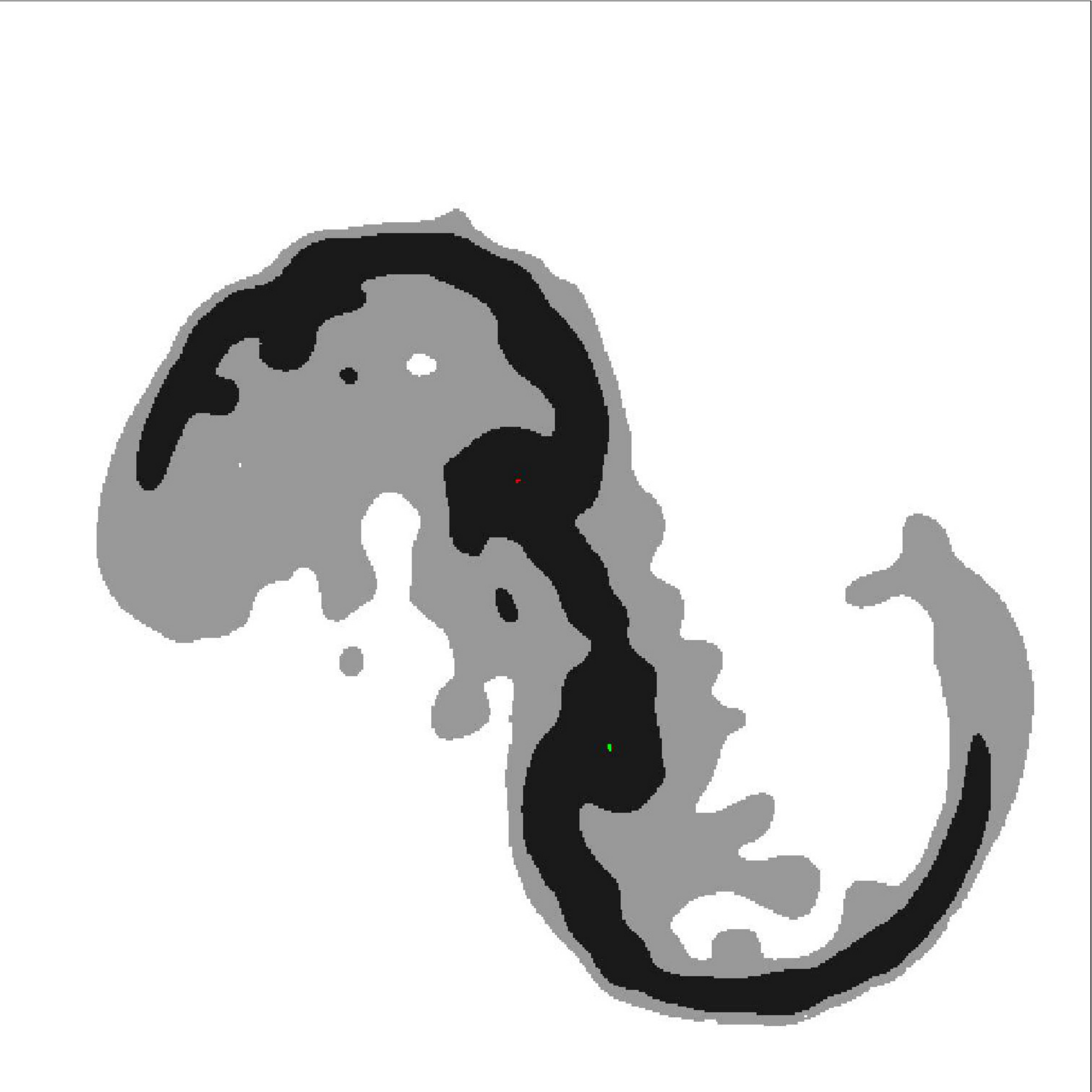}}

\subfloat[]{\label{fig:boxing3}\includegraphics[width=0.4\textwidth]{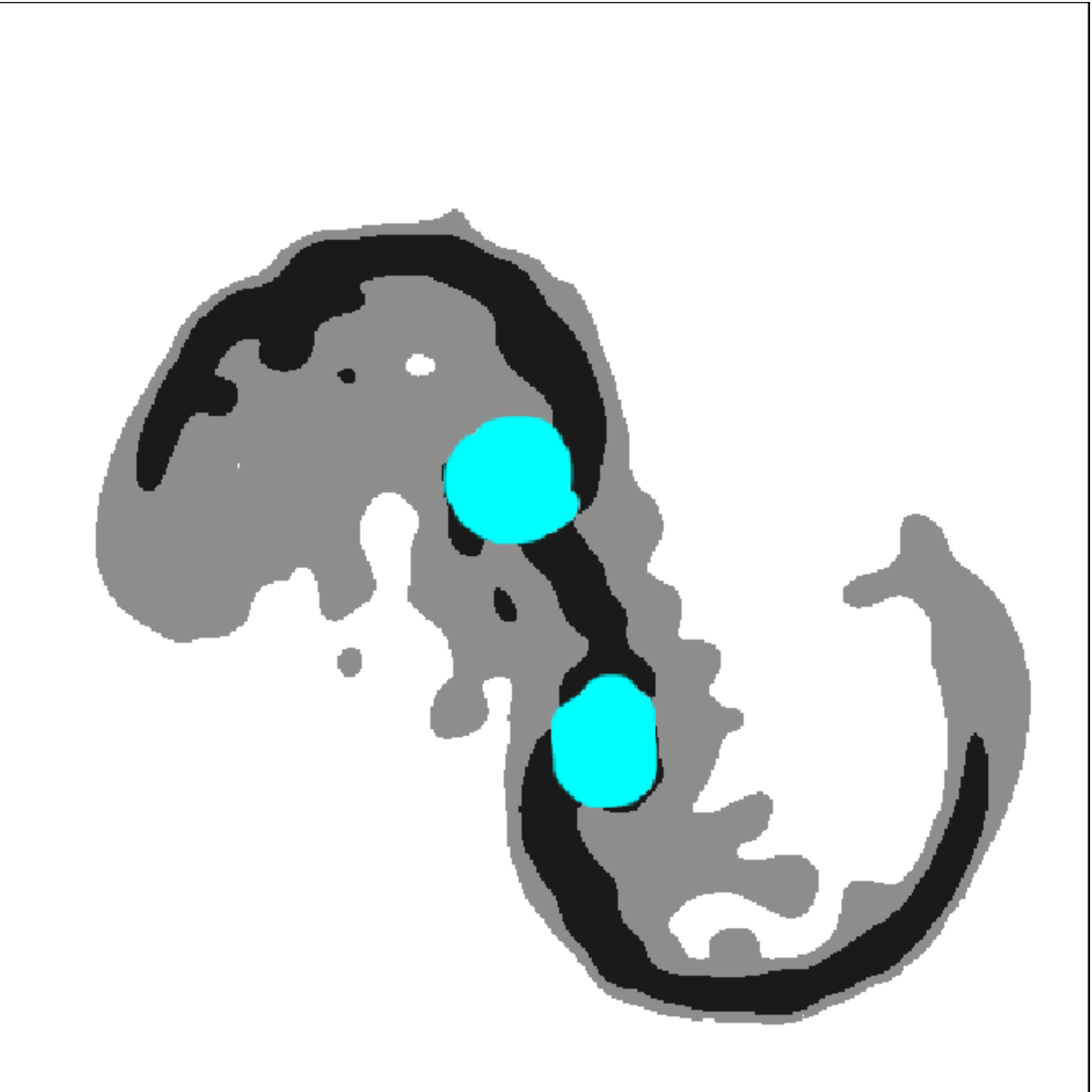}}
\subfloat[]{\label{fig:boxing4}\includegraphics[width=0.4\textwidth]{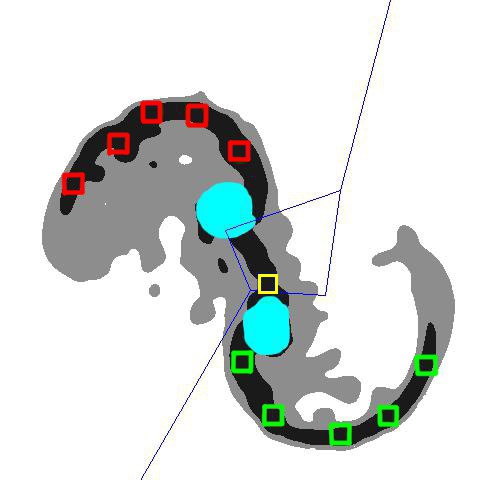}}

\caption{Semi-automated box selection procedure: (a) Cold gas in a GADGET galaxy merger 
simulation (the fiducial model in test 1, see Table \ref{tab:gadget}). (b) segmentation map for
two surface brightness levels. (c) The centers of the two galaxies are masked by the user to
avoid selecting boxes in these regions. (d) We select 11 random points (boxes) in the remaining 
black region. The blue lines divide the regions for disc 1- (red), disc 2- (green),  and the overlap
(yellow) boxes. }
\label{fig:boxing}
\end{figure*}

Even though Identikit 2 reduces the human subjective influence in finding the best match, it still depends on selection of box positions. If we move the boxes in X, Y , or V directions, or change their size or number, we select  different regions of phase space. As a result, test particles populate them differently, leading to different scores. In Identikit one positions these boxes manually using an interactive interface. In order to explore the effect of box-selection and also to reduce the human  interference, we developed a simple semi-automated routine for box selection. In this routine, boxes are selected akin prescribed user-selection in Identikit. Figure \ref{fig:boxing} describes our semi-automated routine.

Our automated box selection routine places the boxes on the tidal features and the bridges of the merger system. The algorithm attempts to place the boxes in the same style as they are placed manually in \cite{Barnes:2011kb}. In this routine, first the segmentation maps of the system with different surface brightness levels are made. The surface brightness levels are manually adjusted to include faint tidal features in the segmentation map. In the next step, the user manually masks the central regions of the galaxies within the segmentation map, where he/she wants to avoid putting boxes. As explained earlier, Identikit reproduces the large scale tidal tails of the interacting galaxies but not the self-gravitating details. Boxes must not cover the centers of the galaxies. Finally, a user-defined number of boxes are randomly placed in the remaining regions of the segmentation map (Figure \ref{fig:boxing}). The velocity of these boxes are determined by calculating the mass-(light-)averaged velocity of the cold gas (stars) inside the box. While there is still human influence in this routine (i.e. adjusting the surface brightness limit for segmentation map and masking the centers of galaxies), it is more automated than the original process in Identikit 2 and can be used for exploring the random effects of box selection. It should be mentioned that this routine only selects boxes of the same size, but \cite{Barnes:2011kb} often uses varying box sizes. 

We have chosen the size of the boxes to be comparable to the spatial resolution of the data available. Here, we assume that the GADGET simulated galaxies are at the same distance as typical nearby major merger systems (e.g. the Mice galaxies). The HI data of nearby interacting galaxies have been obtained using VLA D and C arrays (e.g. \citealt{Hibbard:1996doa}). The typical angular resolution of these observations is $\sim20"-30"$ which is equivalent to $\sim 8-12$ kpc. For GADGET simulations of cold gas, we select the size of the boxes to be $\sim 6-8$ kpc. The optical H$\alpha$ kinematics data obtained by many current instruments usually have a higher resolution. For instance, the diameter of the fibers in SparsePak on the WIYN telescope is $\sim 6"$, equivalent to $\sim 3$ kpc on a system like the Mice (\citealt{Bershady:2004gp}). So, when testing the young stars in our simulations, we selected the boxes to be $\sim 3$ kpc.

The number of boxes also affect the scores because we select different phase space regions to be populated. For each test on GADGET simulations described in \S \ref{sec:hyrodynamical_simulations_vs_identikit}, the number of boxes is fixed. For tests on cold gas, we used a simple method to determine this number. This method is the outcome of our initial trials. First, we obtain the maximum number of boxes that fit inside the allowed regions (black regions in Figure \ref{fig:boxing}). Then, we take 2/3 of this as the number of boxes. As a result, if the tidal tails are larger we select more boxes. The logic behind is that when the tidal features are more extended we have a stronger constraint. The range of the number of boxes in the tests with cold gas is $\sim 8-16$. For tests on young stars, the initial trials indicated that using 10 boxes result in better convergence. We used 10 boxes in these tests. We describe convergence in \S \ref{sec:results}. The number boxes used in each test is given in Table \ref{tab:gadget}.

\subsection{Calculating Uncertainties in Best-Fit Parameters}
\label{sec:calculating_uncertainties_in_best_fit_parameters}

The uncertainty of the best-fit model is not given by Identikit itself. The results of our initial tests with Identikit showed that positioning the boxes is the largest source of uncertainty in the calculated scores. In order to get a measure of the uncertainty of the scores, we move the boxes over the tidal tails and bridges of the data galaxy multiple times. More precisely, we use the semi-automated routine 10-100 times. Given each set of boxes, we measure scores for all Identikit models. The average and standard deviation of scores at each point in the parameter space is calculated. The model with the maximum average score is the best-fit model, and the models for which the average score is within 1, 2, 3, and 4 standard deviations from that of the best fit model are within $1 \sigma$, $2 \sigma$, $3 \sigma$ , and $4 \sigma$ contours in Figure \ref{fig:scoremap}, where $\sigma$ is the standard deviation of scores at the best-fit model.

\begin{figure}
\label{fig:number_of_runs1}\includegraphics[width=0.47\textwidth]{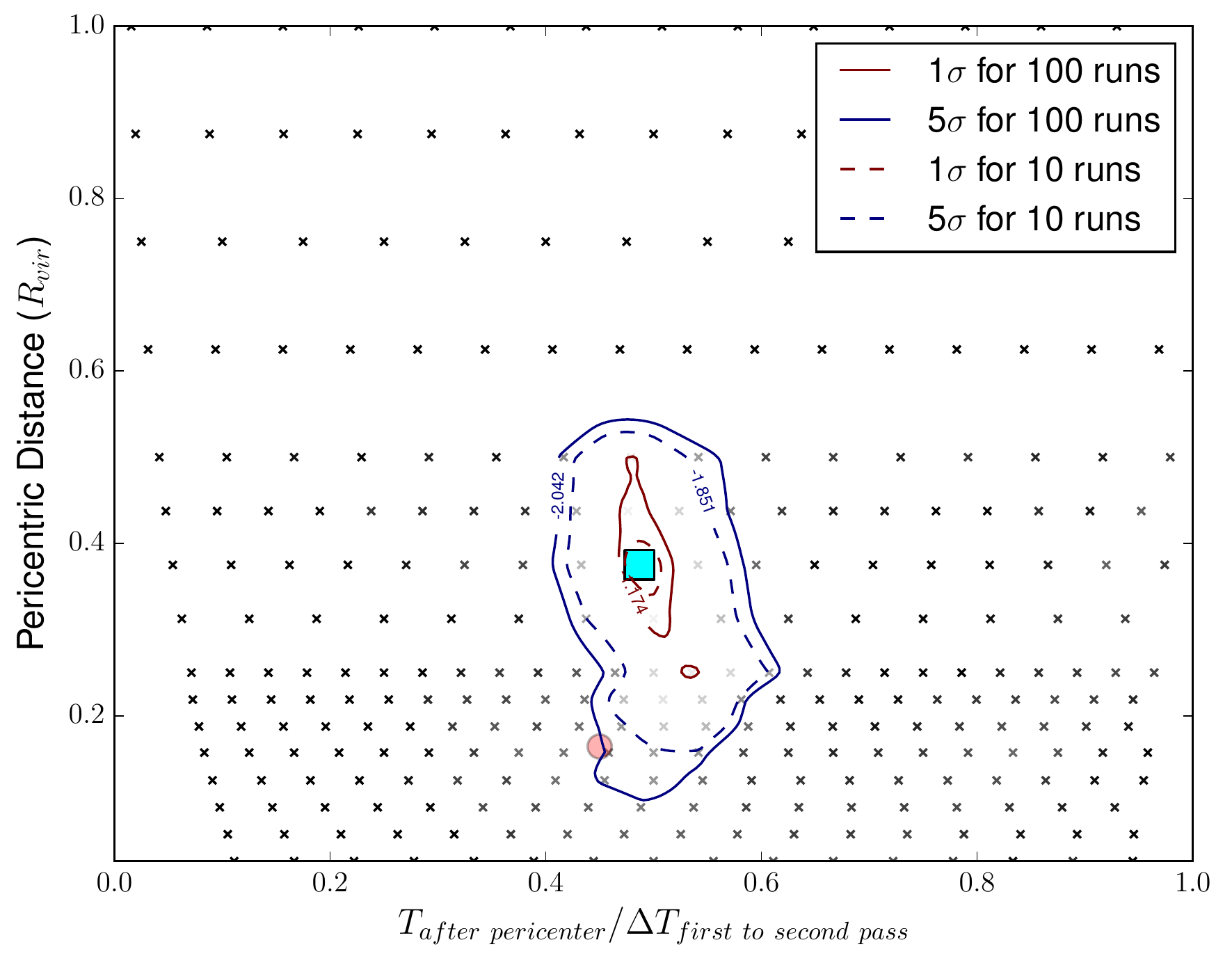}
\caption{The score map with 1$\sigma$ and 5$\sigma$ contours obtained via 
matching the Identikit to the cold gaseous component of the fiducial GADGET 
simulation. The solid contours are calculated by 100 runs and the
dashed contours are calculated by 10 runs.
The 1 $\sigma$ and 5$\sigma$ contours do not change much
by increasing the number of runs by 1 order of magnitude. 
We see the same trend after 20 runs for tests on young stars.
Throughout this work we calculate scores 
10 times for cold gas tests and 20 times for young star
tests.}
\label{fig:number_of_runs}
\end{figure}

Because calculating the score for a full grid of models is a computationally expensive, we need to restrict the number of times in which we move the box positions. To determine this number, we track how the $1\sigma$ and $5\sigma$ level contours change as we increase the number of realization of box positions (runs). We find at what approximate number these contours stop changing significantly with more runs. As can be seen in Figure \ref{fig:number_of_runs}, the contours do not change much after 10 runs in the tests on cold gas. So, we we run these tests 10 times. For the tests on young stars we see the same trend after 20 runs, so we run them 20 times.

\section{Hydrodynamical Simulations Vs. Identikit}
\label{sec:hyrodynamical_simulations_vs_identikit}

In addition to the random uncertainties described above, we can estimate the systematic biases in the best-fit model using an independent set of hydrodynamical simulations. We use the snapshots of the independent simulated systems as mocked data. Once we find the best-fit model, we can compare its parameters with the correct merger parameters of the simulation. 

\subsection{GADGET Simulations} 
\label{sec:gadget_simulations}

We use smoothed particle hydrodynamical (SPH) GADGET-2 simulations of equal-mass disc-disc galaxy merger as input data (\citealt{Cox:2006kb} and \citealt{Cox:2008jj}). The isolated galaxies are Sbc type galaxies with stellar and cold gaseous disc components and stellar bulge component, embedded in a massive dark matter halo. Each simulation follows a galaxy merger in a box of $\sim (200$kpc$)^3$. The spatial resolution is 100 pc, and the particle mass is $\sim 1.0-7.1\times10^5 M_{\odot}$. Each galaxy is made of 1,000,000 dark matter particles, 300,000 each of stellar disc and collisional gas, and 100,000 bulge particles, so there is a total 3.4 million particles in each simulation. The snapshots are available in 0.05 Gyr time steps. The simulations include radiative cooling, a density-dependent star formation recipe that reproduce Kennicutt-Schmidt relation (\citealt{KennicuttJr:1998ki}), and a model to incorporate the effect of supernovae feedback and stellar winds. Even though feedback from active galactic nuclei (\citealt{2008ApJ...676...33D}) is not included, these simulations capture more physical processes than simple test-particle Identikit models, and they are more realistic.

The mass models of isolated galaxies in GADGET simulations are different from Identikit models.  \cite{Cox:2006kb} and \cite{Barnes:2009fh} describe the mass models of the GADGET simulations and Identikit, respectively. The halo concentration parameter is different between the two models ($c=11$ in GADGET simulations and $c=4$ in Identikit). The length scale that is used to match the two simulations is the scale radius of the NFW profile (\citealt{Navarro:1996ce}). The stellar and gaseous disc scale length in the GADGET simulations are 4 kpc and 12 kpc respectively, which are 0.24 and 0.72 of the scale radius of the halo.  In Identikit, isolated galaxies have a disc scale length that is $1/3$ of the halo scale radius. We expect to see biases in the initial conditions of the matched simulations as a result of different mass models. However, in real world, the mass models of the interacting galaxies also differ from that of Identikit, and testing Identikit against a simulation with a different mass model helps us better understand the sensitivity of its results on the selected mass model for Identikit.

\subsection{Model Details}
\label{sec:model_details}

We test Identikit against GADGET simulations described in Table \ref{tab:gadget}. These models are selected to test the capability of Identikit in modeling systems with varying time since pericenter, viewing angles, eccentricity, pericentric distance, and initial disc orientations. We obtained these simulation via private communication with T.J. Cox, and the distribution of their initial parameters was not designed for the purpose of our tests. The fiducial GADGET simulation used in tests 1-6 is an equal-mass galaxy merger with parabolic orbit. The pericentric distance is equal to 11 kpc. In this simulation, both of the galaxies are prograde. In test 01 and test 02 we look at cold gas and young stars in a face-on snapshot midway between the first and the second passage. In most of other tests, all other parameters but one is similar to tests 01 and 02. We investigate the isolated effect of varying each parameter on the systematic and random uncertainties of the best-fit. We examine cold gas (tests 01, 05, 07, 10, 12, and 14) vs. young stars (tests 02, 06, 08, 11, 13, and 15), merger stage (tests 03 and 04), viewing angle (tests 05 and 06), eccentricity (tests 07, 08, and 09), pericentric distance (test 09), and disk orientation (tests 10-15).

\begin{table*}
\centering
\begin{tabular}{cccccccccc}
	\hline
	Sim			&	Test		&	component	 	& 	pericentric			& eccentricity	& prograde 			& 	$(i_1,\omega_1)$	&	view 					& time	$^*$		 				& number 		\\
	ID			&	ID			&	(gas/			&	 distance	  		&				&  vs. 				&	$(i_2,\omega_2)$	&	(to orbital			&$\Delta T/T_{\textrm{first}}$ 		& of			\\  
				&	 			&	young			&						&				& retrograde		&						&	angular				& $_{\textrm{to second}}$			& boxes		\\
				&	 			&	stars)			&						&				& 					&						& 	momentum)	& $_{\textrm{pass}}$				& 				\\ \hline \hline
	Sbc201a	&	01 			&	cold gas				&	$11 kpc $ 			& 1.0 			&prograde			& 	$(0,0)$				& face-on	&  0.45								& 11		\\
				&	(fiducial)	&					&	$=0.1643$			&				&-prograde 		& 	$(30,60)$			&			& 									&			\\
				&				&					&	$R_{vir}$			&				&			 		&						&			& 									&			\\
				&	02			&	young 			&	"					& "				& "					& "						& "		 	&  "									& 10		\\
				&				&	stars			&						&				&		 			&						&			& 									&			\\ 
				&	03			&	cold gas				& 	"		 			& "	 			& "		 			& "						& "			&  0.25 								& 9			\\
				&				&					&						&				&		 			&						&			& 									&			\\ 
				&	04			&	cold gas				& 	"		 			& "	 			& "		 			& "						& "			&  0.75 								& 13		\\
				&				&					&						&				&		 			&						&			& 									&			\\ 
				&	05			&	cold gas				& 	"					& "	 			& "		 			& "						& edge-on	&  0.45 								& 10		\\
				&				&					&						&				&		 			&						&			& 									&			\\ 
				&	06			&	young 			& 	"		 			& "	 			& "		 			& "						& "			&  "	 								& 10		\\
				&				&	stars			&						&				&		 			&						&			& 									&			\\ \hline
	Sbc212		&	07			&	cold gas				&	$11 kpc $ 			& 0.9 			& prograde 			& 	$(0,0)$				& face-on 	& 0.50								& 14		\\
				&				&					&	$=0.1643$			&				&-prograde 		& 	$(30,60)$			&			& 									&			\\
				&				&					&	$R_{vir}$			&				&		 			&						&			& 									&			\\  
				&	08			&	young 			&	"		 			& "				& "		 			& "						& "		 	& "									& 10		\\
				&				&	stars			&	"					&				&		 			&						&			& 									&			\\ \hline 
	Sbc214		&	09			&	cold gas				&	$44 kpc$  			& 0.8 			& prograde			& 	$(0,0)$				& face-on 	& 0.50								& 16		\\
				&				&					&	$=0.6572$			&				&-prograde 		& 	$(30,60)$			&			& 									&			\\
				&				&					&	$R_{vir}$			&				&		 			&						&			& 									&			\\ \hline 
	Sbc207		&	11			&	cold gas				& 	$11 kpc $ 			& 1.0 			& polar				& 	$(270,0)$			& face-on 	& 0.50 								& 12		\\
				&				&					&	$=0.1643$			&				&-prograde 		& 	$(30,60)$			&			&									&			\\
				&				&					&	$R_{vir}$			&				&					&						&			&									&			\\
				&	12			&	young 			& 	"					& "				& "					& "						& "		 	& "	 								& 10		\\
				&				&	stars			&						&				&		 			&						&			&									&			\\ \hline 
	Sbc203		&	13			&	cold gas				& 	$11 kpc $ 			& 1.0 			& retrograde		& 	$(180,0)$			& face-on 	& 0.50 								& 8			\\
				&				&					&	$=0.1643$			&				&-retrograde 		& 	$(210,60)$			&			& 									&			\\
				&				&					&	$R_{vir}$			&				&			 		&						&			& 									&			\\
				&	14			&	young 			& 	"					& "	 			& "					& "						& "		 	& "	 								& 10		\\
				&				&	stars			&						&				&		 			&						&			& 									&			\\ \hline
	Sbc202		&	14			&	cold gas				& 	$11 kpc $ 			& 1.0 			& retrograde		& 	$(180,0)$			& face-on 	& 0.50								& 10		\\
				&				&					&	$=0.1643$			&				&-prograde 		& 	$(30,60)$			&			& 									&			\\
				&				&					&	$R_{vir}$			&				&		 			&						&			& 									&			\\
				&	15			&	young 			& 	"					& "	 			& "					& "						& "			& "									& 10		\\
				&				&	stars			&						&				&		 			&						&			& 									&			\\ \hline
	\hline
\end{tabular}
\caption{The GADGET simulations used to test Identikit modeling in this work. 
$^*$ Time is the fraction of time after first passage to the time between the first and the second passage.}
\label{tab:gadget}
\end{table*}

We can put a prior constraints on some of the encounter parameters by selecting interacting galaxies at certain stages of the encounter. Identikit 2 can only model merger systems with separate galaxies with distinct cores. In these systems we can estimate the merger mass ratio by measuring its light ratio. In addition, we are looking for galaxy merger systems that have strong tidal features which are the most pronounced after the first passage. We select test GADGET simulations at times long enough after their first pass and before their second pass, such that they have separate nuclei and strong tidal tails. 

\begin{figure*}
\centering
\includegraphics[width=0.90\textwidth]{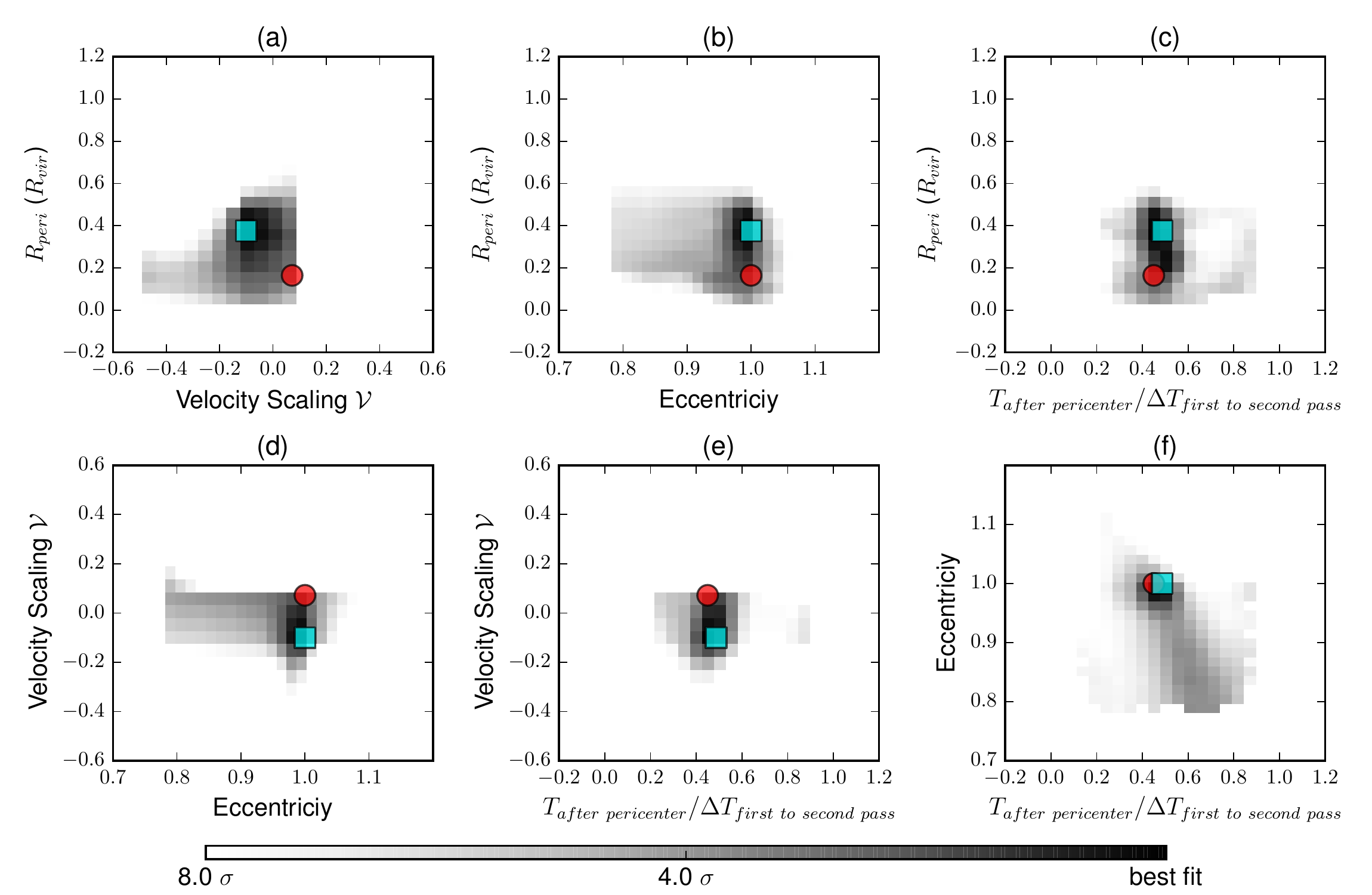}
\caption{The average score maps of Identikit models 
matching to our fiducial GADGET simulation (Sbc201a, test 01). These are 2+1-D slices of the score map 
for varying (a) velocity scaling and pericentric separation, (b) eccentricity and pericentric separation,
(c) time and pericentric separation, (d) eccentricity and velocity scaling, 
(e) time and velocity scaling, and (f) time and eccentricity. These slices are taken
at the best fit point which is shown by the cyan square. The color scale
goes from best-fit (black) to 8$\sigma$ from best fit (white). The red 
circles show the correct encounter parameters of the GADGET simulation. }
\label{fig:bestfitcut01}
\end{figure*}

The hydrodynamical simulations have separate gaseous and stellar components, with distinguishable young and old stellar populations. For Identikit modeling, line of sight velocity information is required. This informations can be obtained from cold gas (e.g. HI 21 cm emission), ionized HII regions (e.g. H$\alpha$ emission),  molecular clouds (i.e. CO emission), or stars (absorption). Cold neutral gas is usually more extended than the stellar population in galaxies. Resultingly, it shows more pronounced tidal features in galaxy mergers, and so it is expected to be easier to model with Identikit. However, obtaining high resolution HI data is observationally more expensive than H$\alpha$ emission line maps. In this work, we compare the result of Identikit modeling using cold gas (representing HI emission), vs.  young stars (representing H$\alpha$ emission from HII regions). 

\section{Results}
\label{sec:results}

The primary output of the routine we described in \S \ref{sec:method} is a score map for each test in Table \ref{tab:gadget}.  The score maps are 4+1-dimensional  spaces (eccentricity, pericentric distance , time, velocity scale + score). As described in \S \ref{sec:identikit}, each point in the score map refers to a member of the Identikit model library with a particular velocity scaling  $\mathcal{V}$. The viewing angle, orientation of discs, and length scaling $\mathcal{L}$ is set by calculating the score for each Identikit model. The uncertainty of the scores at  every point in the score map is obtained by measuring the scores 10-100 times, each with a new randomly selected set of boxes (\S \ref{sec:box_selection}), and calculating the average and standard deviation of scores at each point. The best matched Identikit model is expected to be that with the highest average score. This model is called the best-fit model and its parameters are the best-fit parameters. Models with average scores within $n$ standard deviations of that of the best-fit model are considered to be within $n\ \sigma$ of best-fit model. In order to visualize the variation of scores in the 4+1-D score map, we can look at its 2+1 dimensional slices at the position of best-fit model and see the shape of the region with high score in different directions. As there are 6 ways to choose 2 parameters out of 4, we show six 2+1-D score maps in Figure \ref{fig:bestfitcut01}. The extent of the $n$-$\sigma$ contours in each of these four directions determines the uncertainty of the corresponding parameter.

\begin{figure*}
\includegraphics[width=0.7\textwidth]{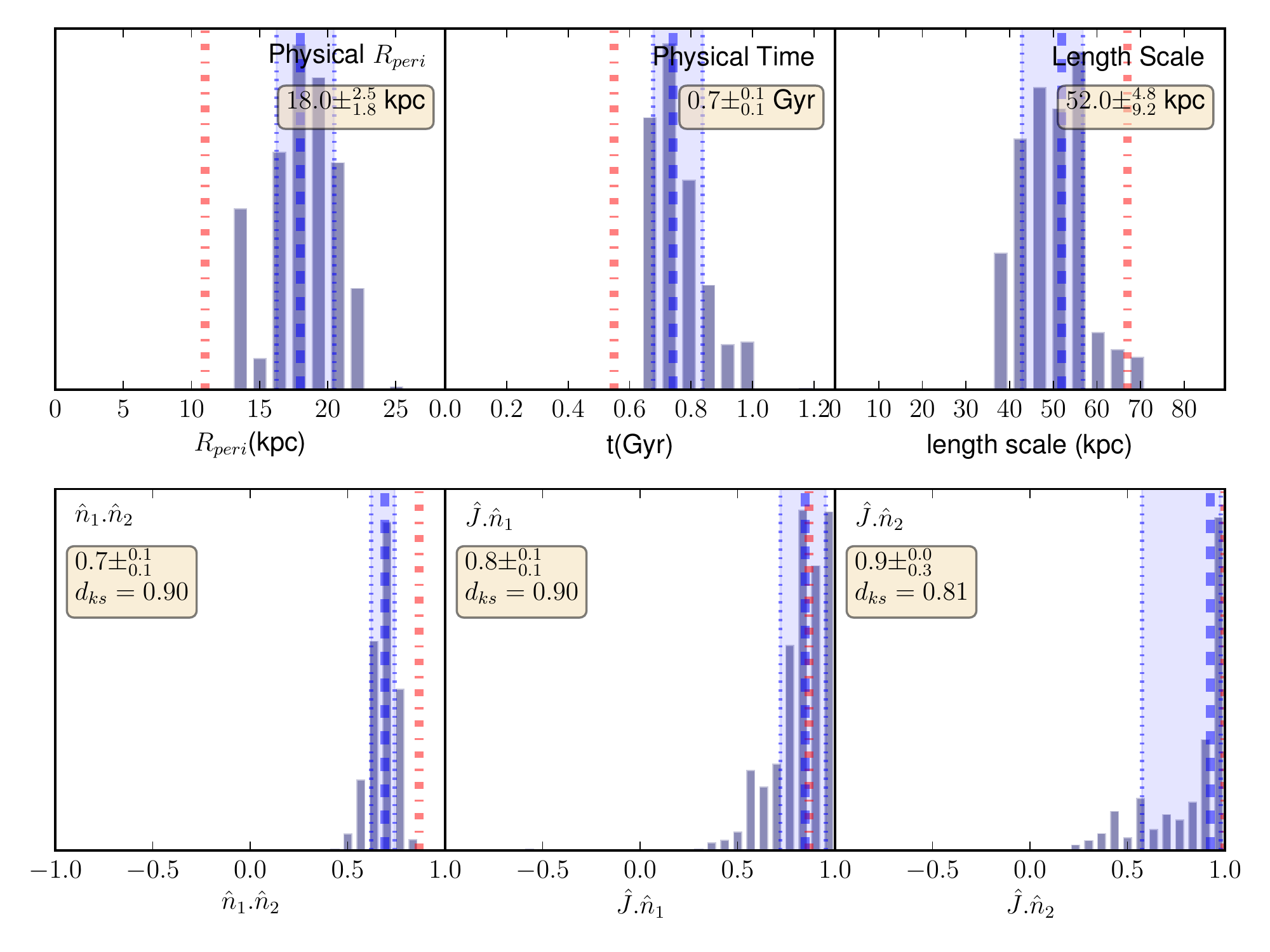}
\caption{The distribution of parameters in models within 1$\sigma$ 
contours in Figure \ref{fig:scoremap}. The blue dashed vertical line shows the
median of the distribution and the red dot-dashed vertical line shows 
the correct value of the parameter in
the GADGET simulations. The shaded areas show the interval between 
the 16 and 84 percentiles.
The median and the limits of the shaded area
 are given in each panel. On the lower plots the KS test 
value against a cosine distribution is written. This test is considered as a good 
convergence because the average of $d_{KS}>0.75$. Tests with $0.75>d_{KS}>0.30$ are grouped
as fairly converged, and tests with $0.30>d_{KS}$ are poorly converged.}
\label{fig:parameterdist01}
\end{figure*}

\begin{figure*}
\subfloat[]{\label{fig:angledist01}\includegraphics[width=0.4\textwidth]{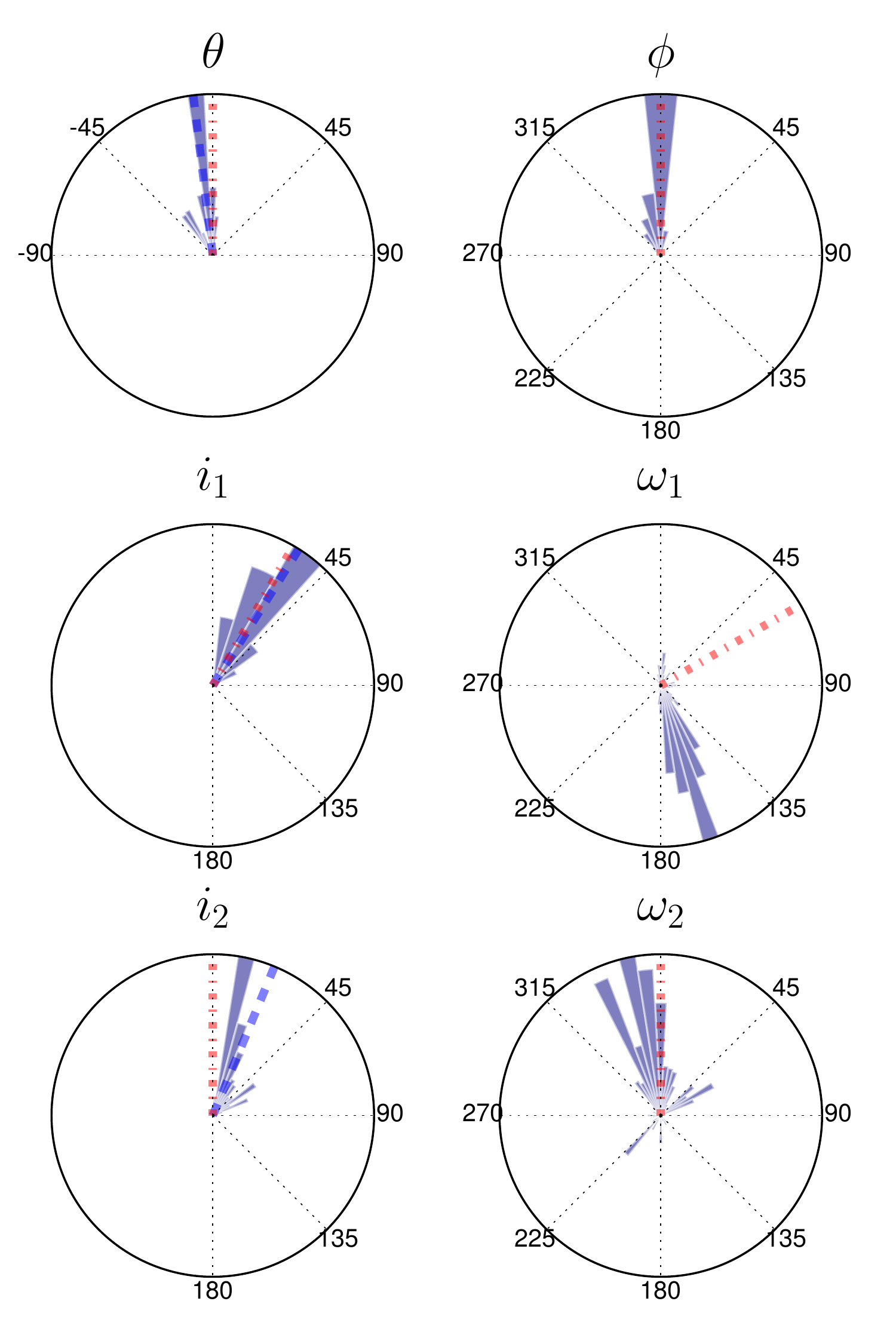}}
\subfloat[]{\label{fig:angledist12}\includegraphics[width=0.4\textwidth]{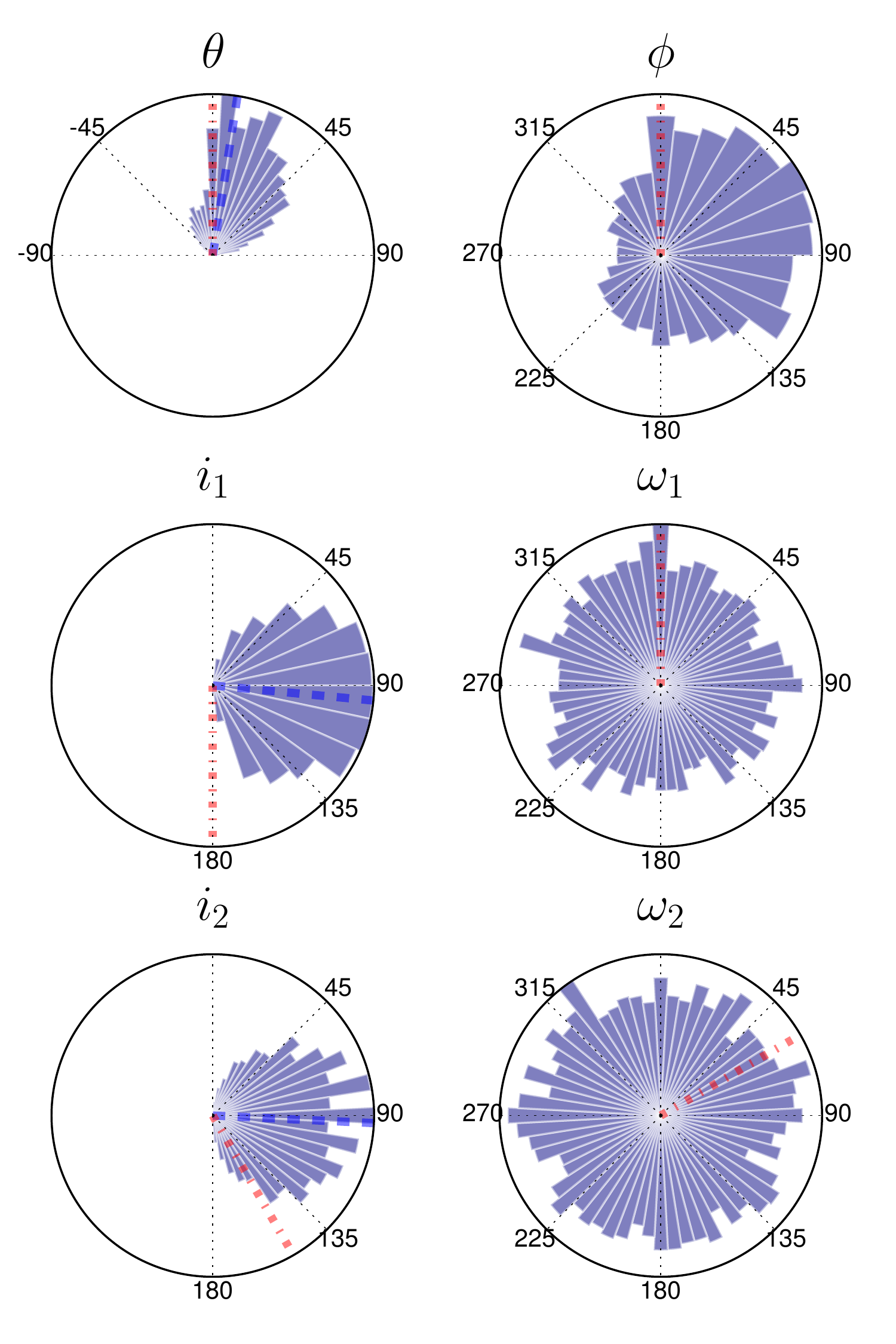}}
\caption{The distribution of the angles in models within 
1$\sigma$ (a) for test 01 which converges well
and (b) for test 13 which does not converge. $\theta$ and $\phi$ are viewing angles, and
$i_d$ and $\omega_d$ are inclinations and argument to 
periapsis of disc d. Blue radial dashed lines show the 
median and the red radial dot-dashed lines show the correct angles. 
Note that in (a) $i_1$ is close to 0 which corresponds to the northern pole of the coordinate system. 
So, the difference between the correct $\omega_1$ (red dot-dashed radial line) and the measured peak represents
 a small angular separation, and the distribution of disc 1 
 orientations is close to the correct value. In (b) the flat distribution of
 histograms indicate that the orientation of discs and viewing 
 angles of the best models are random and one cannot rely
on the answer for other parameters.  We quantify convergence by 
performing a KS test on the distribution of $\hat{n}_i.\hat{J}$. 
The KS test value for (a) and (b) in this figure are 0.87 and 0.21, respectively.
Among the 15 tests in this work we found 7 good, 4 fair, and 4 poor convergences.}
\label{fig:angledist}
\end{figure*}

In order to measure the uncertainty of the parameters of the best-fit model determined inside Identikit and not explicitly tested in the grid of models (i.e. viewing angle, orientation of discs, length scaling  $\mathcal{L}$, and physical time and physical $R_{peri}$ which depend on  $\mathcal{L}$), we study the distribution of model parameters with scores within the 1$\sigma$ contour in Figure \ref{fig:scoremap}. Figure \ref{fig:parameterdist01} shows these distributions for test 01. Here the distribution of physical pericentric distance ($\mathcal{L}\times R_{peri}/R_{vir})$, physical time ($\mathcal{L}/ \mathcal{V}\times T_{simulation}$), length scaling  $\mathcal{L}$, $\hat{n}_1.\hat{n}_2$ (dot product of orientation of the 2 discs), and $\hat{n}_i.\hat{J}$ (dot product of the orientation of disc $i$ and system's orbital angular momentum) are shown. The latter is a quantitive measure for the system being prograde vs. retrograde. When $\hat{n_i}.\hat{J}=1$ we have a prograde disc, when $\hat{n_i}.\hat{J}=-1$ we have a retrograde disc, and when $\hat{n_i}.\hat{J}=0$ we have a polar disc. The correct parameter from GADGET simulation is shown with a dot-dashed red vertical line so one can easily see the bias in the distribution. We use rose plots  to show the distribution of angles in models within 1$\sigma$ and 3$\sigma$ contours. Figure \ref{fig:angledist01} shows the rose plots for test 01. It includes the distribution of altitude ($\theta$) and azimuth ($\phi$) of the viewing direction, along with the  inclination ($i$) and argument to periapsis ($\omega$) of each disc. The correct angles are shown with red dot-dashed radial lines.

In some of the tests the 1 $\sigma$ distribution of initial orientation of discs converges into a narrow range of angles (Figure \ref{fig:angledist01}). In these tests the models close to the best-fit model find close initial disc orientations, and we may infer that these models are small variations of the best-fit model. However, the 1$\sigma$ distribution of initial orientation of discs are wider for some tests, and in some cases these distributions are almost flat (Figure \ref{fig:angledist12}). A wide or flat distribution indicates that the models with similarly high scores have very different initial disc orientations, and there is a significant degeneracy in the best-fit model. In the most extreme case, when all disc orientations are equally likely we would find a cosine distribution in $\hat{n}_1.\hat{n}_2$ and $\hat{n}_i.\hat{J}$ histograms.

We categorize the results of our tests into three groups based on the distribution of $\hat{n}_i.\hat{J}$ for models within 1$\sigma$ of the best-fit model. Tests with relatively narrow peak in  $\hat{n}_i.\hat{J}$ distribution are ``well-converged". Tests in which the peak is considerably broader are ``fairly-converged". Finally, ``poor'' convergence is when we have an almost flat distribution in  $\hat{n}_i.\hat{J}$. In order to quantify this categorization, we perform a KS test of the distribution of $\hat{n}_i.\hat{J}$ against a cosine distribution. A large KS test value means that the distribution is not a cosine function, and a small value means it is similar to a cosine function. When the KS test value ($d_{KS}$) is bigger than 0.75, we label it as good convergence. When it is less that 0.30 we take the result as a poor convergence, and when it is between 0.30 and 0.75 we consider it a fair convergence. Note that KS test value limits are somewhat arbitrarily selected and can be changed to include more or less tests in good and fair groups. However, as can be seen in the best-fit vs. correct parameter plots later in this section, our selection of good, fair, and poor convergence is correlated with the accuracy of the results.

Seven out of 15 tests in Table \ref{tab:gadget}  have good convergence, four have fair convergence, and four are poorly converged. All of the tests on prograde-prograde systems have good or fair convergence. Tests on edge-on systems resulted in good convergence as well. Tests on retrograde and polar systems have either poor convergence or are on the lower end of $d_{KS}$ range for fair convergence. The reason for poor convergence is either because Identikit can not find a configuration that populates the boxes, or it is because the boxes are not restrictive enough on orientation of discs and models with different orientations give similarly good scores. We will discuss this in \S \ref{sec:discussion}.

 \subsection{The Fiducial Test}

Figures \ref{fig:bestfitcut01} and \ref{fig:parameterdist01}  show the score map and distribution of parameters for test 01. This is a test on cold gaseous components of the the fiducial GADGET simulation (Sbc201a). The face-on snapshot is near the middle of the first and the second passages. Eccentricity and pericentric distance of the orbit is 1.0 and 11 kpc, respectively, and both galaxies are prograde (one of them is completely prograde and the other one is tilted by 30 degrees). We identify this test as a good converging test because the average of KS test values for $\hat{n}_i.\hat{J}$ and $\hat{n}_1.\hat{n}_2$ distributions against cosine function is bigger that 0.75.

As can be seen from Figure \ref{fig:bestfitcut01}, our method rules out a significant fraction of parameter space in test 01. The scores of the Identikit models with correct eccentricity (e=1) are significantly higher than the models with incorrect eccentricity, making eccentricity measurement accurate ($\sigma = 0.03$) with no systematic bias. In this case $\sigma$ is set as the half of the increment in eccentricity of Identikit models used (grid size; see Table \ref{tab:Identikit}).The fractional time since pericenter is also well constrained. Fractional time since pericenter is the fraction of current time since pericenter to the total time between the first pass and the second pass. In this work, we chose most of the GADGET models in the mid-way between the first pass and the second pass (i.e. fractional time $\approx$ 0.5). In this particular test the correct fractional time is 0.45. The obtained fractional time from Figure \ref{fig:bestfitcut01} is 0.49 $\pm^{0.15}_{0.09}$. This time is dimensionless, and in order to find the physical time since pericenter we need to scale it using the length and velocity scalings ($\mathcal{L}/\mathcal{V}$). The velocity scaling is one of the free parameters in our tests. However, as described in \S \ref{sec:identikit}, when we lock the centers of the galaxies length scaling is determined by the viewing angle. Any bias in the measured viewing angle leads to a bias in length scaling. The distribution of length scalings of the models within 1$\sigma$ contour are shown on the top-right panel of Figure \ref{fig:parameterdist01}. One can see that the length scaling is biased. However, the velocity scaling of models within the 1$\sigma$  contour is also biased in the opposite direction (see Figures \ref{fig:bestfitcut01}a, \ref{fig:bestfitcut01}d, and \ref{fig:bestfitcut01}e). The distribution of physical time since pericenter for these models is not as biased. (See top-middle panel of Figure \ref{fig:parameterdist01}.) The measured physical time since pericenter is 0.74$\pm^{0.10}_{0.07}$ Gyrs, which is within 3$\sigma$ of the correct answer (0.55 Gyrs). The orientation of discs with respect to orbital angular momentum is shown in $\hat{n_i}.\hat{J}$ panels in Figure \ref{fig:parameterdist01}. For both discs the distribution correctly indicates prograde disc orientation. The correct answer is also recovered for viewing angles $\theta$ and $\phi$ in Figure \ref{fig:angledist01}.

Pericentric separation is not as well constrained. The black regions on the score maps of Figure \ref{fig:bestfitcut01} is broad in the direction of pericentric distance ($0.38\ R_{vir}\pm^{0.16}_{0.16}R_{vir}$). Nevertheless, the correct fractional pericentric separation $0.16\ R_{vir}$, is within 1.5$\sigma$ of the best-fit value. In addition, the physical pericentric distance is obtained by scaling the dimensionless $R_{peri}$ with the length scaling parameter $\mathcal{L}$. The distribution of physical pericentric distances is shown in the top-left panel of Figure \ref{fig:parameterdist01}. The median of this distribution is 18.1$\pm^{2.5}_{1.8}$ kpc. This is also within 3$\sigma$ of the correct value, with a relatively smaller value for $\sigma$. 

We present the result of other tests by examining the parameters adjusted in the GADGET simulations in Table \ref{tab:gadget}.

\subsection{Time Since Pericenter}

In tests 03 and 04 (see Table \ref{tab:gadget}) we use the exact same GADGET model used in test 01, but observed at earlier and later fractional times respectively. Fractional time is the time since the first passage divided by the total time between the first and the second passages, so it is a measure of the merger stage. Test 03 is fairly converged with average $d_{KS}$ = 0.55. In this test, the correct fractional time is 0.25, which is equivalent to 0.30 Gyr after the first pass. The obtained fractional time for this model is 0.22$\pm^{0.17}_{0.09}$ and the measured physical time since pericenter is 0.36$\pm^{0.97}_{0.24}$ Gyr. Test 04 is well converged. In this test, the correct fractional time is 0.75, which is equivalent to 0.85 Gyrs after pericenter. The best-fit fractional time is 0.86$\pm^{0.01}_{0.01}$ and the obtained best-fit physical time is 1.23$\pm^{0.15}_{0.14}$ Gyr. Fractional time is biased especially in the late stage test (test 04); however, it is encouraging that we recover the fractional time within 10\% of the correct value.

\begin{figure}
\subfloat[]{\label{fig:cor_vs_bestfit_reltime}\includegraphics[width=0.45\textwidth]{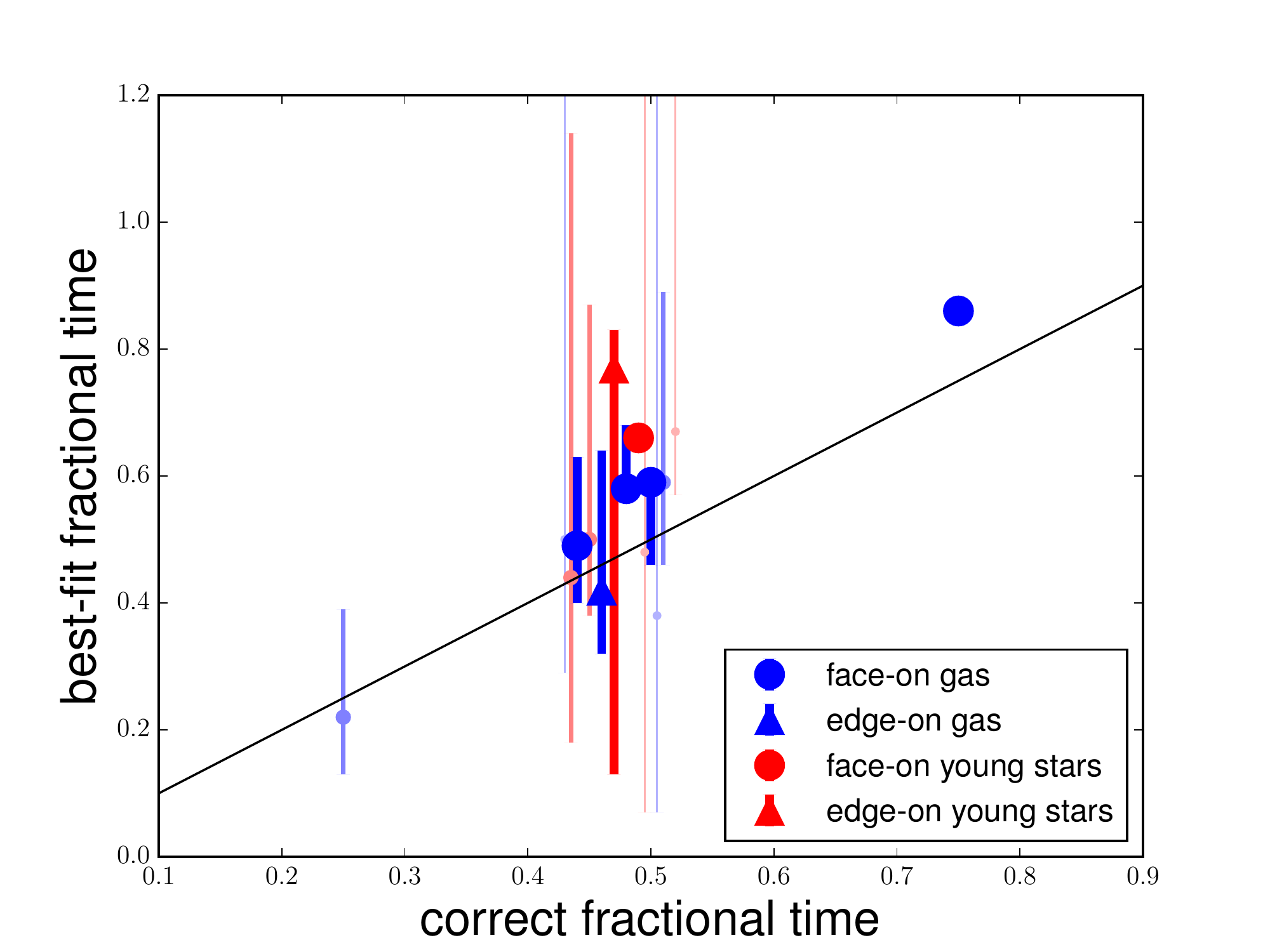}}\\
\subfloat[]{\label{fig:cor_vs_bestfit_phystime}\includegraphics[width=0.45\textwidth]{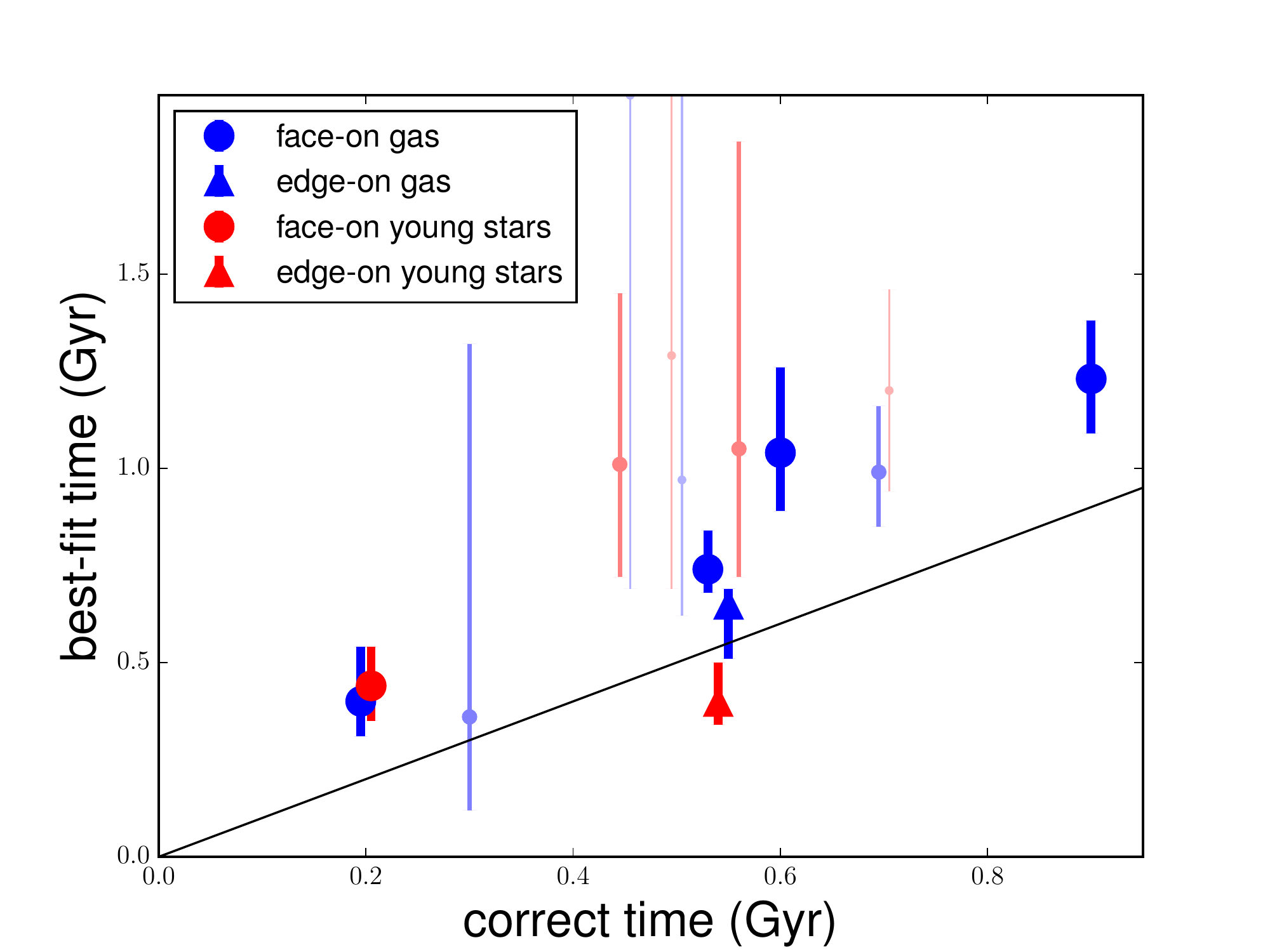}}
\caption{Correct vs. best-fit values of (a) fractional time and (b) physical time since pericenter 
for all the tests in this work. The blue markers are test on cold gaseous component,
and red markers are tests on young stars. The face-on tests are shown with circles and the edge-on tests are
shown with triangles. Tests with good, fair, and poor convergence are shown with big, medium and small markers, respectively.}
\label{fig:cor_vs_bestfit_time}
\end{figure}

In the rest of the tests we use snapshots of the GADGET simulations that are near the middle of the first and the second passages (fractional time $\approx$ 0.5). Figure \ref{fig:cor_vs_bestfit_time} shows the best-fit vs. correct values of fractional time and physical time since pericenter. In this figure we use different marker sizes, line thicknesses, and color darknesses for tests with different convergences to emphasize good and fair tests. We do the same for all best-fit vs. correct plots in this work. We only study tests with good and fair convergence. Parameters in tests with poor convergence usually have large uncertainties and are not reliable measurements. They are plotted, however, for the sake of comparison. The tests with good and fair convergence are following the black line with a scatter of $\approx$ 0.2 in fractional time, and the best-fit fractional times are mostly within 1 $\sigma$ of the correct value. 

The physical time, which is obtained with a combination of length and velocity scalings ($\mathcal{L}/\mathcal{V}$), is more clearly biased toward later times in tests on face-on systems (see Figure \ref{fig:cor_vs_bestfit_phystime}). The average bias for face-on tests is $\approx$ 0.3 Gyrs. This systematic effect can be corrected when we apply our method on real data. It is also worth noting that this bias is different for good and fair tests; the physical time in fair tests are more overestimated. The results of tests on edge-on systems seem to be closer to the correct physical time, indicating that the extra informations in velocity gradients of an edge-on viewed system helps to better constrain the merger stage.

\subsection{Eccentricity}

\begin{figure}
\includegraphics[width=0.45\textwidth]{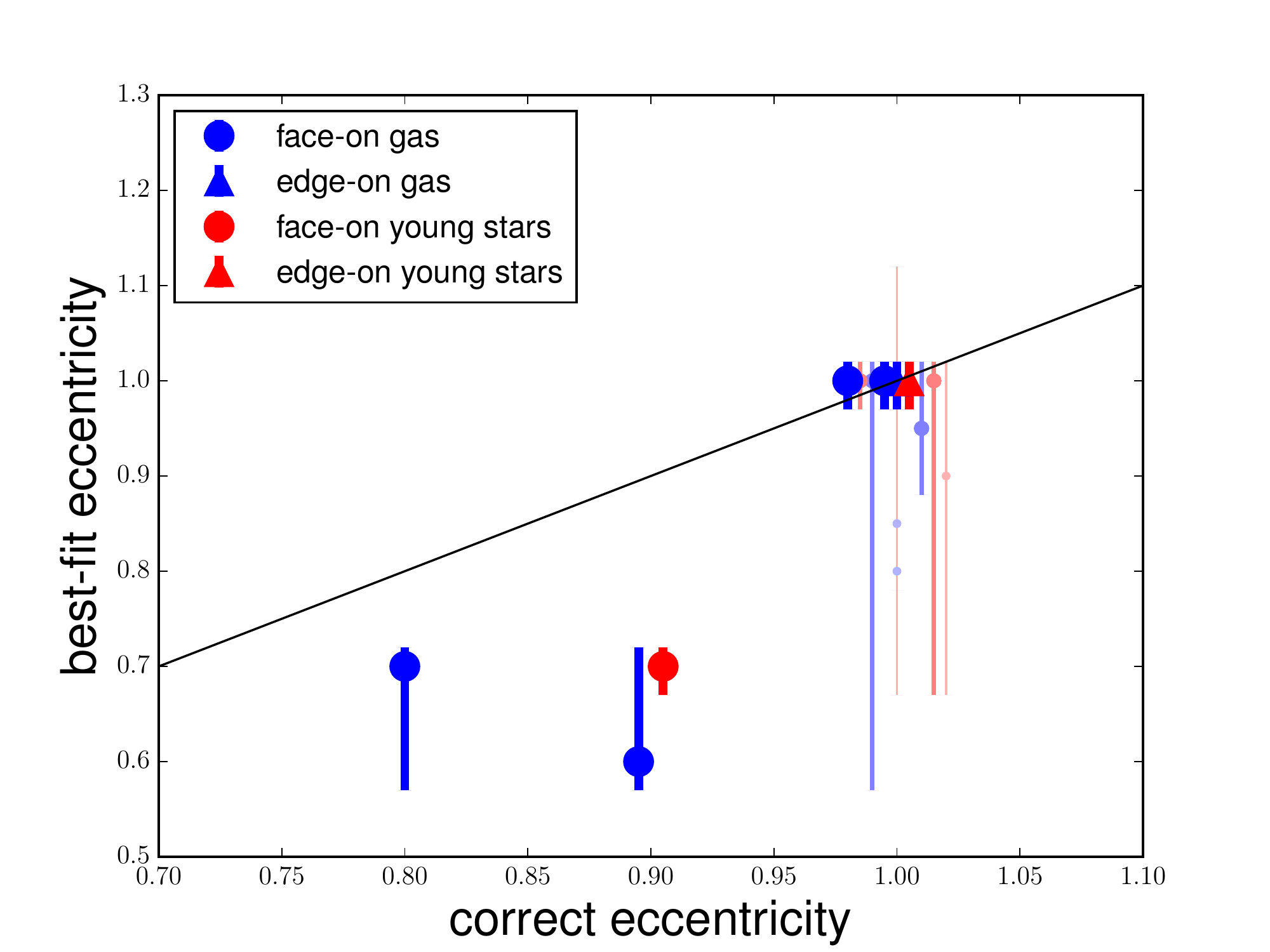}
\caption{(a) Correct vs. best-fit values of eccentricity for all of the tests in this work. 
Symbols and colors are as in Figure \ref{fig:cor_vs_bestfit_time}.
The correct values have been artificially scattered a little to make the points and error bars visible.}
\label{fig:cor_vs_bestfit_eccent}
\end{figure}

Most of the GADGET simulations we used in this work have eccentricity=1.0 (i.e. parabolic orbit). This corresponds to Keplerian orbits with zero energy, which is preferred if galaxies and dark matter halos they live in start approaching each other from stationary and gravitationally unbound origins. In all of the good and fair tests with eccentricity equal to 1.0, we recover the parabolic orbit of the interacting galaxies within 1$\sigma$. In Figure \ref{fig:cor_vs_bestfit_eccent} which shows the measured eccentricity vs. correct eccentricity all of parabolic tests are on the black line. Their correct eccentricity is artificially slightly randomized to improve visibility. 

All of the tests on elliptical orbits result in good convergence. In these tests, we recover the elliptical nature of the orbit, though the eccentricity is underestimated. In test 07 on cold gas with correct eccentricity of 0.9, the measured eccentricity is $0.60\pm^{0.15}_{0.05}$. Test 08 on the young stars of the same simulation result in a slightly better eccentricity ($0.7\pm^{0.05}_{0.05}$). In test 09 in which the correct eccentricity is 0.8 and the pericentric distance is larger than other tests, we also obtain an elliptical orbit, with 0.1 less eccentricity than the correct value ($e=0.70\pm^{0.13}_{0.03}$). 

\subsection{Pericentric Separation}

\begin{figure}
\subfloat[]{\label{fig:cor_vs_bestfit_relperi}\includegraphics[width=0.45\textwidth]{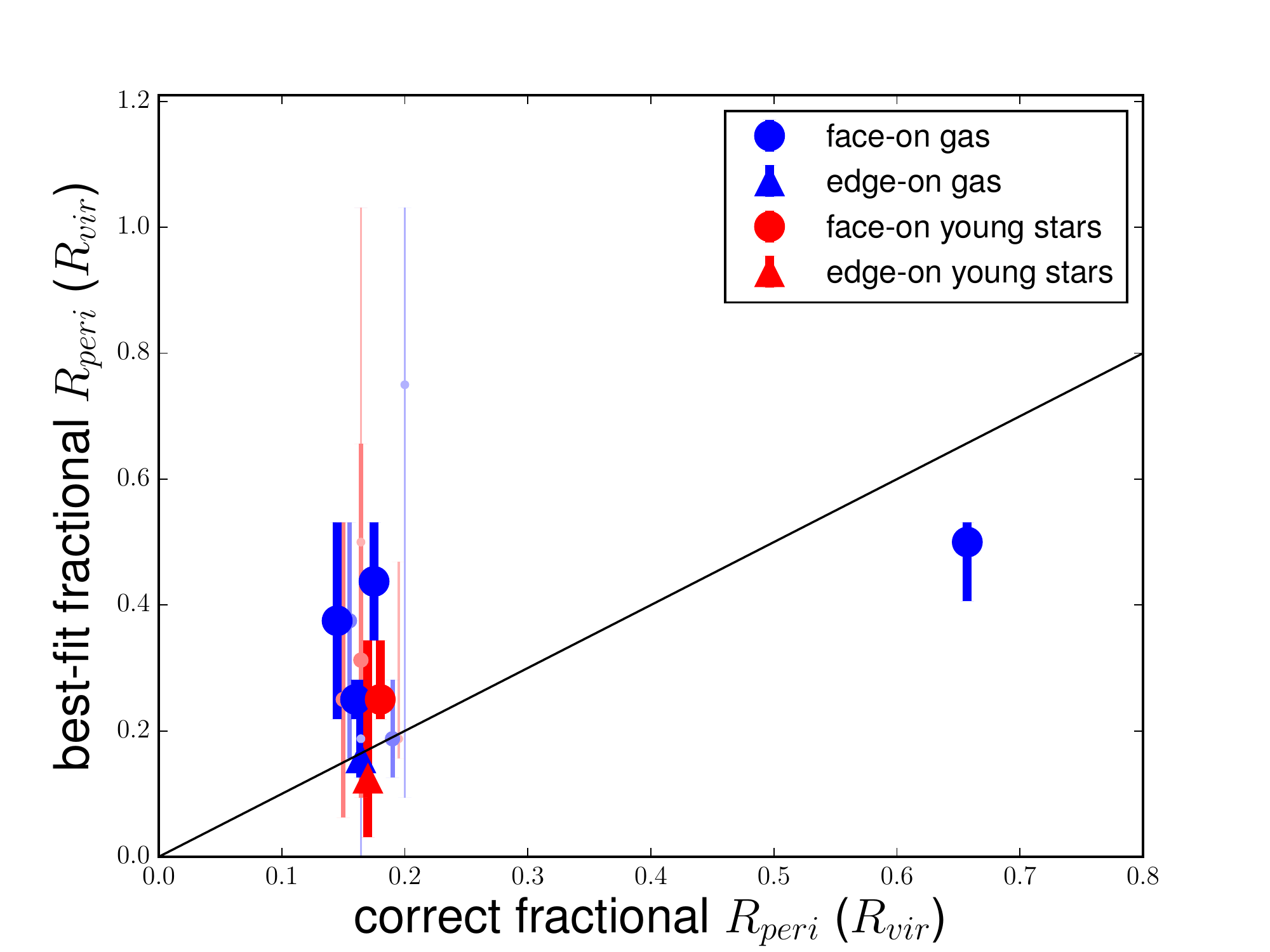}}\\
\subfloat[]{\label{fig:cor_vs_bestfit_physperi}\includegraphics[width=0.45\textwidth]{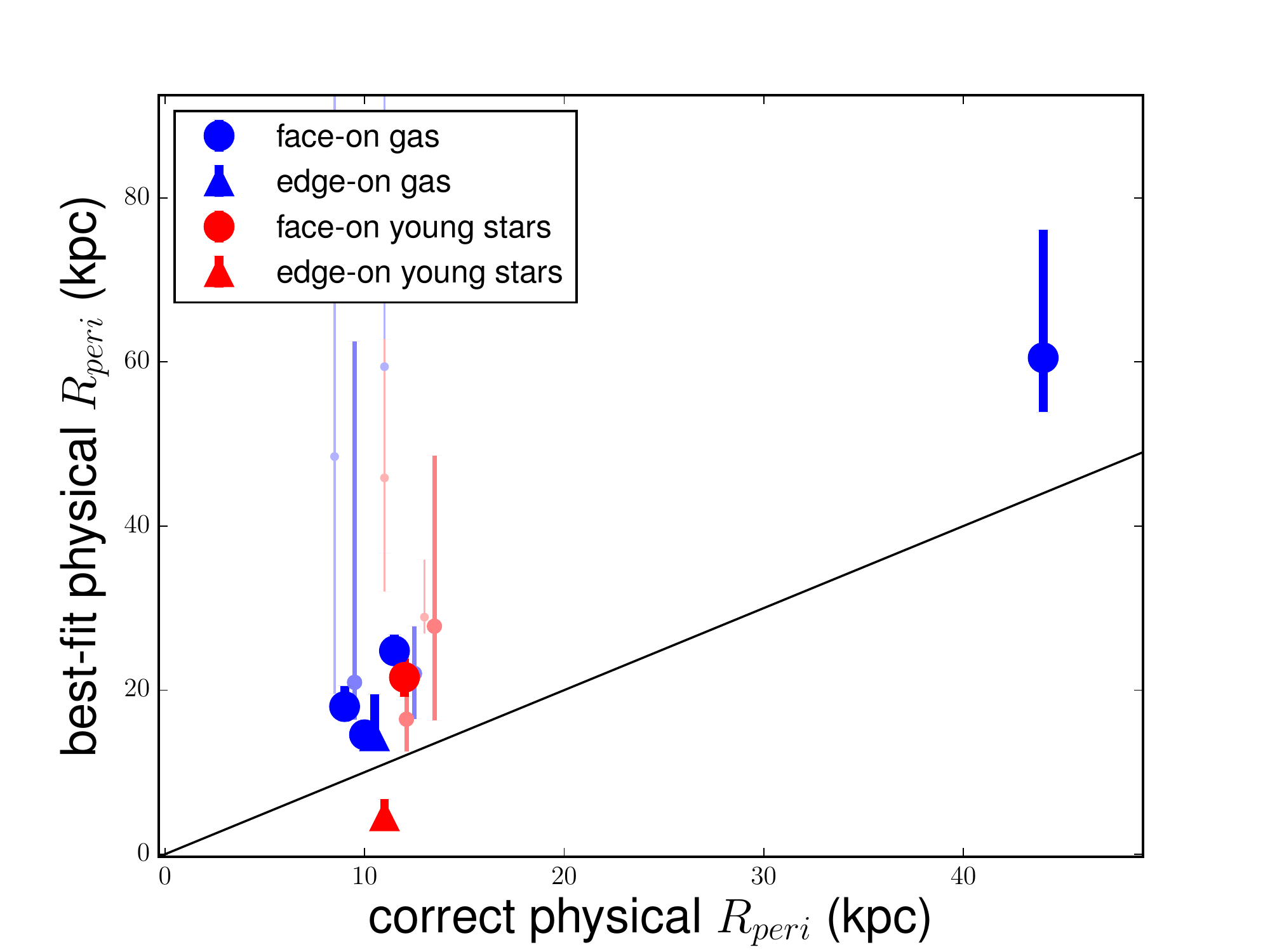}}
\caption{Correct vs. best-fit values of (a) dimensionless and (b) physical
 pericentric distance. 
 Symbols and colors are as in Figure \ref{fig:cor_vs_bestfit_time}.
The correct values have been artificially scattered a little to make the points and error bars visible.}
\label{fig:cor_vs_bestfit_peri}
\end{figure}

Figure \ref{fig:cor_vs_bestfit_peri} shows the best-fit vs. correct answer for both the dimensionless ($R_{peri}/R_{vir}$) and physical pericentric distance. Identikit measures the pericentric separation in units of the virial radius of isolated galaxies which is the dimensionless pericentric distance. In order to find the physical pericentric distance one has to scale dimensionless pericentric distance with the length scaling, $\mathcal{L}$. All but one of the GADGET simulations we tested have the same input $R_{peri}$ (11 kpc).

Fractional pericentric distance is within 3$\sigma$ of the correct value in all of well, and fairly converged tests (Figure \ref{fig:cor_vs_bestfit_relperi}). However, when scaled with the length scaling $\mathcal{L}$ to obtain the physical pericentric distance, all but one (test 06) move to slightly above the black line within 2$\sigma$ (Figure \ref{fig:cor_vs_bestfit_physperi}). Physical $R_{peri}$ in the face-on good and fair tests are overestimated by an average of $\approx$ 30\%. This systematic can be corrected when we apply our method on real data. The results of edge-on tests, however, are closer to correct pericentric distance, which confirms that it is easier to model edge-on interacting discs.


\subsection{Viewing Angle}

\begin{figure}
\includegraphics[width=0.45\textwidth]{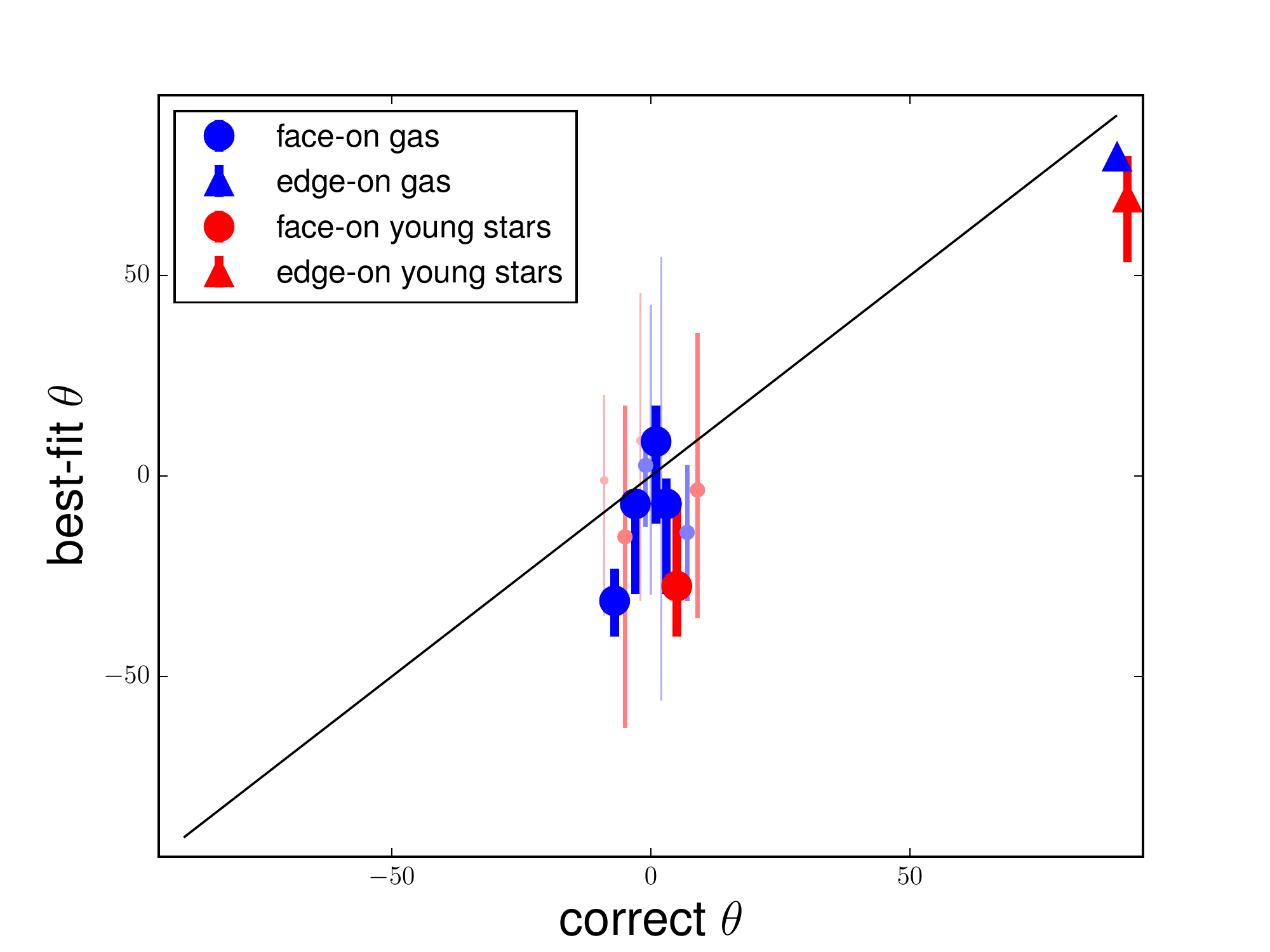}
\caption{Correct vs. best-fit values of altitude of viewing angle. 
Symbols and colors are as in Figure \ref{fig:cor_vs_bestfit_time}.
The correct values have been artificially scattered a little to make the points and error bars visible. The error bar of the 
test on the edge-on cold gas system is too small to be visible. }
\label{fig:cor_vs_bestfit_theta}
\end{figure}

Figure \ref{fig:cor_vs_bestfit_theta} shows the correct vs. meausred altitude in viewing angle. Most of the tests are face-on (Correct $\theta = 0^{\circ}$). Good and fair tests are scattered in the range from $-30^{\circ}$ to $+10^{\circ}$ degrees. It can be seen that there is no major difference between tests on gas and young stars when it comes to best-fit viewing direction of face-on systems.

Two of the tests (05 and 06) are on edge-on systems (Correct $\theta = 90$). Both tests converge and the answers for $\theta$ of the viewing angle are within $10^{\circ}$ and $20^{\circ}$ for the cold gas and young star tests, respectively. We can see in Figures \ref{fig:cor_vs_bestfit_time} and \ref{fig:cor_vs_bestfit_peri} that edge-on tests result in better best-fit physical time and pericentric distance than most of face-on tests. This suggests that in edge-on systems we have better constraints to model mergers stage and $R_{peri}$. In a merger of edge-on discs, the velocity difference in tidal streams are more visible in the line of sight direction. The better visiblity of velocity variations puts a stronger constraint on length and velocity scalings which results in better constraints on dependent parameters.

\subsection{Initial Orientation of Discs}

\begin{figure}
\subfloat[]{\label{fig:cor_vs_bestfit_Jni}\includegraphics[width=0.45\textwidth]{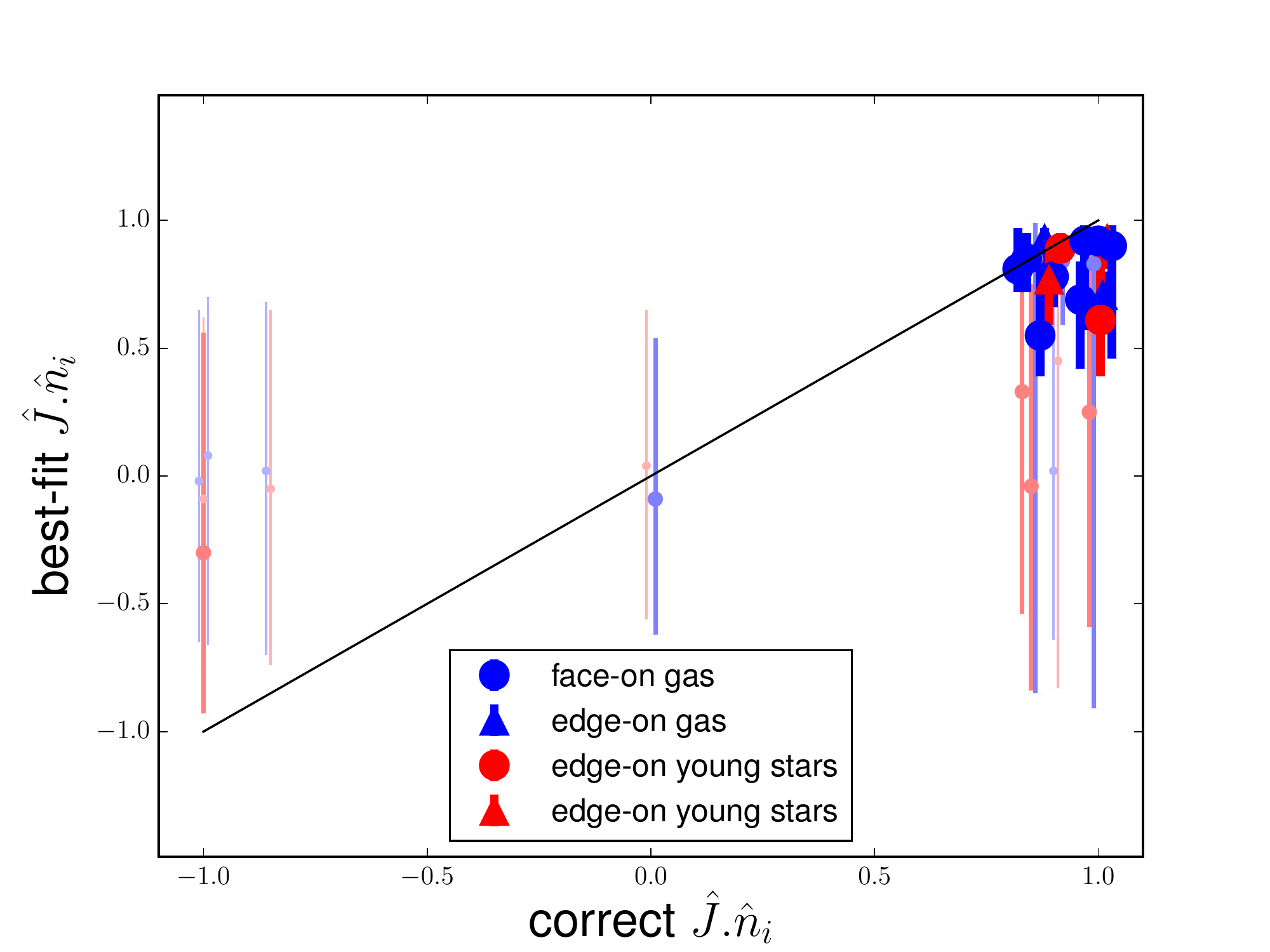}}\\
\subfloat[]{\label{fig:cor_vs_bestfit_n1n2}\includegraphics[width=0.45\textwidth]{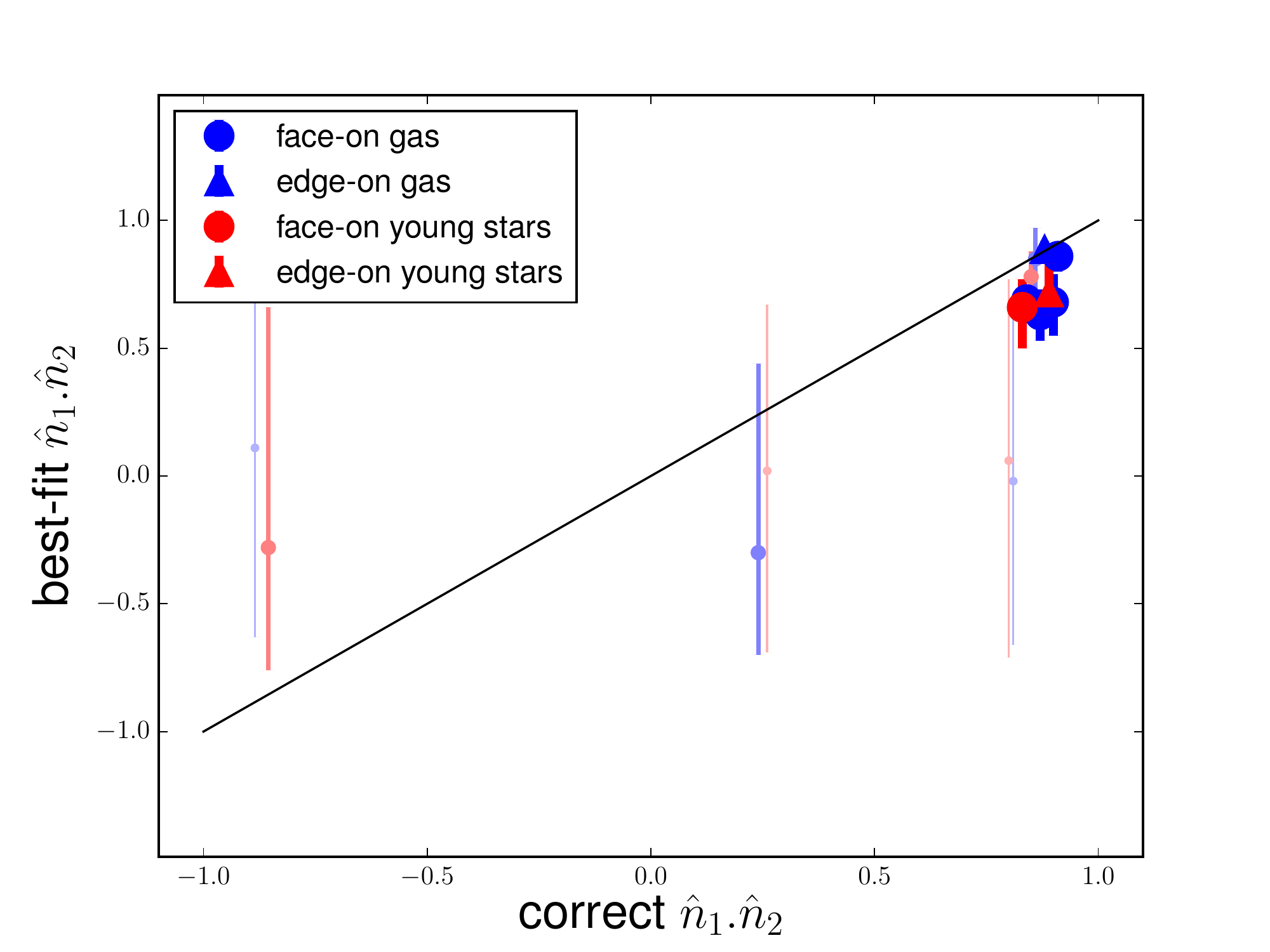}}
\caption{Correct vs. best-fit values of (a) $\hat{n_i}.\hat{J}$, and 
(b) $\hat{n_1}.\hat{n_2}$. 
Symbols and colors are as in Figure \ref{fig:cor_vs_bestfit_time}.
The correct values have been artificially scattered a little to make the points and error bars visible. }
\label{fig:cor_vs_bestfit_orientaiton}
\end{figure}

Figure \ref{fig:cor_vs_bestfit_orientaiton} shows the correct vs. obtained values for orientation of discs. We use $\hat{n_i}.\hat{J}$ to show the orientation of discs with respect to the orbital angular momentum of the galaxy merger. $\hat{n_i}$ is the unit vector of the orientation of disc i and $\hat{J}$ is the unit vector of angular momentum of the orbit. So, when $\hat{n_i}.\hat{J}=1$ we have a prograde disc, when $\hat{n_i}.\hat{J}=-1$ we have a retrograde disc, and when $\hat{n_i}.\hat{J}=0$ we have a polar disc. Values in between these reflect systems that are more or less prograde, polar, or retrograde. We also use $\hat{n_1}.\hat{n_2}$ to show the orientation of discs with respect to each other.

Nine out of 15 tests are on prograde-prograde galaxy mergers. All but two of them (tests 02 and 03) resulted in good convergence. Tests 02 and 03 are also fairly converged, and the average $d_{KS}$ for them is on the higher end of the range for fair convergence. The measured  $\hat{n_i}.\hat{J}$ and $\hat{n_1}.\hat{n_2}$ for all of these tests are within 0.3 of the correct answer ($\approx 1.0$).

The remaining six tests are on prograde-polar (tests 10 and 11), retrograde-retrograde (tests 12, 13), and prograde-retrograde (tests 14,15) orbits. All but two of them (tests 10 and 15) result in poor convergence. Tests 10 and 15 have fair convergence, though they are also on the lower end of the range of $d_{KS}$ for fair convergence. The measured uncertainties of $\hat{n_i}.\hat{J}$ in these fair tests are bigger than 0.5 meaning that the orientation of discs are not well constrained in them either. We expected poorer match for retrograde systems as the tidal features are less pronounced. However in original Identikit paper (\citealt{Barnes:2009fh}) some artificial merger models with retrograde discs were successfully reproduced by visual matching in the interactive interface of Identikit. Our failure in modeling retrograde and polar systems can be due to different reasons including the fact that the gaseous tidal tails in retrograde systems have an altered morphology compared to the collisionless stars, and that the tidal tails are even less strong when we study young stars. We will discuss this in more details in \S \ref{sec:discussion}.

 
 \begin{table*}
\centering
\begin{tabular}{ccccccc}
\hline
	Test 				 & 	component 	&  	number	  	& 	convergence			&	best-fit							& 	best-fit 					& best-fit time 						\\
	ID					&	explored		&	of matching	&	quality					&	pericentric  						&	eccentricity				&	$\Delta T/T_{\textrm{first to}}$ 	\\  
						&	 				&	 runs			& 	(KS test value)		&	distance		  					&								& 	${_{\textrm{second pass}}}^*$	 \\ \hline \hline
01						&	cold gas			&	100			&	good					&	$18.0 kpc $						&	1.00						&	0.49									\\
						&					&					&	(0.87)					&	$(\pm^{2.5}_{1.8} kpc)$		&	$(\pm^{0.03}_{0.03} )$	&	$(\pm^{0.15}_{0.09} )$				\\
						&					&					&							&	$=0.38\ R_{vir}$ 				&								&	$=0.74$								\\
						&					&					&							&	$(\pm^{0.16}_{0.16}R_{vir})$	&								&	$(\pm^{0.10}_{0.07} )$Gyr			\\
02						&	young stars	&	100			&	fair						&	$16.5 kpc $						&	1.00						&	0.50									\\
						&					&					&	(0.55)					&	$(\pm^{8.9}_{3.9} kpc)$		&	$(\pm^{0.03}_{0.03} )$	&	$(\pm^{0.37}_{0.12} )$				\\
						&					&					&							&	$=0.25\ R_{vir}$ 				&								&	$=1.05$								\\
						&					&					&							&	$(\pm^{0.28}_{0.19}R_{vir})$	&								&	$(\pm^{0.79}_{0.33} )$Gyr			\\
03						&	cold gas			&	10				&	fair						&	$21.0 kpc $						&	1.00						&	0.22									\\
						&					&					&	(0.69)					&	$(\pm^{41.6}_{4.6} kpc)$		&	$(\pm^{0.03}_{0.43} )$	&	$(\pm^{0.17}_{0.09} )$				\\
						&					&					&							&	$=0.38\ R_{vir}$ 				&								&	$=0.36$								\\
						&					&					&							&	$(\pm^{0.16}_{0.22}R_{vir})$	&								&	$(\pm^{0.97}_{0.24} )$Gyr			\\
04						&	cold gas			&	10				&	good					&	$14.6 kpc $						&	1.00						&	0.86									\\
						&					&					&	(0.86)					&	$(\pm^{0.1}_{0.4} kpc)$		&	$(\pm^{0.03}_{0.03} )$	&	$(\pm^{0.01}_{0.01} )$				\\
						&					&					&							&	$=0.25\ R_{vir}$ 				&								&	$=1.23$								\\
						&					&					&							&	$(\pm^{0.03}_{0.03}R_{vir})$	&								&	$(\pm^{0.15}_{0.14} )$Gyr			\\
05						&	cold gas			&	100			&	good					&	$14.4 kpc $						&	1.00						&	0.42									\\
						&					&					&	(0.96)					&	$(\pm^{5.0}_{0.8} kpc)$		&	$(\pm^{0.03}_{0.03} )$	&	$(\pm^{0.22}_{0.09} )$				\\
						&					&					&							&	$=0.16\ R_{vir}$ 				&								&	$=0.65$								\\
						&					&					&							&	$(\pm^{0.12}_{0.03}R_{vir})$	&								&	$(\pm^{0.04}_{0.14} )$Gyr			\\ 
06						&	young stars	&	10				&	good					&	$4.7 kpc $						&	1.00						&	0.77									\\
						&					&					&	(0.86)					&	$(\pm^{2.0}_{1.3} kpc)$		&	$(\pm^{0.03}_{0.03} )$	&	$(\pm^{0.06}_{0.65} )$				\\
						&					&					&							&	$=0.12\ R_{vir}$ 				&								&	$=0.40$								\\
						&					&					&							&	$(\pm^{0.22}_{0.09}R_{vir})$	&								&	$(\pm^{0.11}_{0.05} )$Gyr			\\ \hline 
07						&	cold gas			&	10				&	good					&	$24.8 kpc $						&	0.60						&	0.58									\\
						&					&					&	(0.82)					&	$(\pm^{2.0}_{0.4} kpc)$		&	$(\pm^{0.13}_{0.03} )$	&	$(\pm^{0.09}_{0.01} )$				\\
						&					&					&							&	$=0.44\ R_{vir}$ 				&								&	$=0.40$								\\
						&					&					&							&	$(\pm^{0.09}_{0.09}R_{vir})$	&								&	$(\pm^{0.14}_{0.09} )$Gyr			\\
08						&	young stars	&	20				&	good					&	$21.6 kpc $						&	0.70						&	0.66									\\
						&					&					&	(0.79)					&	$(\pm^{2.2}_{2.3} kpc)$		&	$(\pm^{0.03}_{0.03} )$	&	$(\pm^{0.01}_{0.01} )$				\\
						&					&					&							&	$=0.25\ R_{vir}$ 				&								&	$=0.44$								\\
						&					&					&							&	$(\pm^{0.09}_{0.03}R_{vir})$	&								&	$(\pm^{0.10}_{0.09} )$Gyr			\\  \hline 
09						&	cold gas			&	10				&	good					&	$60.5 kpc $						&	0.70						&	0.59									\\
						&					&					&	(0.88)					&	$(\pm^{15.6}_{6.6} kpc)$		&	$(\pm^{0.03}_{0.13} )$	&	$(\pm^{0.01}_{0.13} )$				\\
						&					&					&							&	$=0.50\ R_{vir}$ 				&								&	$=1.04$								\\
						&					&					&							&	$(\pm^{0.03}_{0.09}R_{vir})$	&								&	$(\pm^{0.22}_{0.15} )$Gyr			\\ \hline 
10						&	cold gas			&	10				&	fair						&	$22.0 kpc $						&	0.95						&	0.59									\\
						&					&					&	(0.42)					&	$(\pm^{5.8}_{5.6} kpc)$		&	$(\pm^{0.08}_{0.08} )$	&	$(\pm^{0.30}_{0.13} )$				\\
						&					&					&							&	$=0.19\ R_{vir}$ 				&								&	$=0.99$								\\
						&					&					&							&	$(\pm^{0.09}_{0.06}R_{vir})$	&								&	$(\pm^{0.17}_{0.13} )$Gyr			\\ 
11						&	young stars	&	20				&	poor					&	$28.9 kpc $						&	0.90						&	0.67									\\
						&					&					&	(0.24)					&	$(\pm^{7.0}_{2.0} kpc)$		&	$(\pm^{0.13}_{0.23} )$	&	$(\pm^{0.68}_{0.09} )$				\\
						&					&					&							&	$=0.19\ R_{vir}$ 				&								&	$=1.20$								\\
						&					&					&							&	$(\pm^{0.28}_{0.03}R_{vir})$	&								&	$(\pm^{0.26}_{0.26} )$Gyr			\\ \hline
12						&	cold gas			&	10				&	poor					&	$59.4 kpc $						&	0.80						&	0.38									\\
						&					&					&	(0.17)					&	$(\pm^{34.1}_{22.8} kpc)$		&	$(\pm^{0.23}_{0.13} )$	&	$(\pm^{1.63}_{0.31} )$				\\
						&					&					&							&	$=0.75\ R_{vir}$ 				&								&	$=0.97$								\\
						&					&					&							&	$(\pm^{0.28}_{0.66}R_{vir})$	&								&	$(\pm^{3.37}_{0.35} )$Gyr			\\ 
13						&	young stars	&	20				&	poor					&	$45.9 kpc $						&	1.00						&	0.48									\\
						&					&					&	(0.21)					&	$(\pm^{16.9}_{13.8} kpc)$		&	$(\pm^{0.13}_{0.33} )$	&	$(\pm^{2.01}_{0.41} )$				\\
						&					&					&							&	$=0.50\ R_{vir}$ 				&								&	$=1.29$								\\
						&					&					&							&	$(\pm^{0.53}_{0.34}R_{vir})$	&								&	$(\pm^{1.42}_{0.60} )$Gyr			\\  \hline
14						&	cold gas			&	10				&	poor					&	$48.5 kpc $						&	0.85						&	0.50									\\
						&					&					&	(0.17)					&	$(\pm^{96.2}_{28.9} kpc)$		&	$(\pm^{0.28}_{0.08} )$	&	$(\pm^{20.71}_{0.21} )$			\\
						&					&					&							&	$=0.19\ R_{vir}$ 				&								&	$=1.96$								\\
						&					&					&							&	$(\pm^{0.84}_{0.19}R_{vir})$	&								&	$(\pm^{5.06}_{1.28} )$Gyr			\\ 
15						&	young stars	&	20				&	fair						&	$27.8 kpc $						&	1.00						&	0.44									\\
						&					&					&	(0.32)					&	$(\pm^{20.8}_{11.5} kpc)$		&	$(\pm^{0.03}_{0.33} )$	&	$(\pm^{0.70}_{0.26} )$				\\
						&					&					&							&	$=0.31\ R_{vir}$ 				&								&	$=1.01$								\\
						&					&					&							&	$(\pm^{0.34}_{0.22}R_{vir})$	&								&	$(\pm^{0.44}_{0.29} )$Gyr			\\  \hline
	\hline
\end{tabular}
\caption{List of measured parameters in each test.}
\label{tab:results}
\end{table*}


\section{Discussion}
\label{sec:discussion}
Our method results in good and fair convergence for all of the prograde-prograde tests; nevertheless all but two of the retrograde and polar tests have poor convergence. In the converging tests, we can rule out a large fraction of the parameter space, albeit with systematic offsets from the input parameters.

\subsection{Parameters in Converged Tests}

The random uncertainties are obtained by calculating how the scores are affected when we move the boxes around on tidal tails. In our early work on Identikit, we noticed that box positioning had the biggest effect on scores. Nevertheless, there may be other sources of random error (e.g. error from noise in the data) that we did not take into account in our calculations. The uncertainties we measured are lower limits to the real uncertainties. 

Time since pericenter (both physical and fractional) is the best constrained parameter in tests with good and fair convergence. This is an important parameter as it tells us whether the interacting galaxies are in early or late merger stage. We selected most of the snapshots to be in the midway between the first and the second pass. In a couple of tests we recovered the time of an early and a late stage galaxy merger (see Figure \ref{fig:cor_vs_bestfit_time}). In tests of edge-on systems we recovered the correct merger time to 1$\sigma$ level. Though, there is an average bias of $\sim$ 0.3 Gyrs toward later physical times in tests on face-on systems. This bias could be due to the different mass models of the two simulations. The difference in the luminous to dark matter scale length affects the development of tidal features and the bridges. A more extended tidal feature may start to fall back into a galaxy earlier, making it look like a model with shorter tidal feature at a later time. Alternatively, this bias may be originating from the different nature of particles in GADGET simulations vs. Identikit. If the later is the main cause of the bias, we can correct for it when this method is applied on real data. With more tests on a variety of mass models we can track the cause of this bias more precisely.

Eccentricity, pericentric distance, viewing angle, and initial orientation of discs are other parameters that are relatively well constrained. The tested GADGET simulations had limited range for these parameters, because we did not make them ourselves. Eight out of 11 good and fair converged tests have parabolic orbit, and for all of them the eccentricity is correctly recovered, within 1$\sigma$. We claim that our method is successful in measuring the eccentricity of parabolic systems. We also recovered the elliptical nature of tests with eccentricity $e < 1$, though the eccentricity is underestimated by $\sim$ 0.1-0.3. In all of the good and fair tests the measured viewing angle is within $30^{\circ}$ of the correct value (See Figure \ref{fig:cor_vs_bestfit_theta}). Fractional pericentric distance is within 3$\sigma$ of the correct value in all of tests with good and fair convergence; though, there is an average bias of $\sim$ 30\% toward larger physical $R_{peri}$ in face-on tests. The pericentric separation is hard to model, even when visual matching is utilized and the sophisticated patterns recognizable to human brain are matched. The cause of the bias may also be the different mass models of the two simulations or because of using test-particles in Identikit. The initial disc orientations are also well constrained in prograde-prograde tests. In these tests the measured $\hat{n_i}.\hat{J}$s and $\hat{n_1}.\hat{n_2}$s are within 0.3 of the correct value.

Cold gas and young stars reveal consistent results. In simulations that we test both cold gas and young stars the measured values for eccentricity, pericentric distance, merger stage, viewing angle, and disc orientations are within 1$\sigma$ of each other. This suggests that dynamical modeling of galaxy mergers can be done using both H$\alpha$ and HI kinematics data. Only in tests on the fiducial GADGET simulation cold gas result in better convergence than young stars (tests 01 and 02). This is because young stars are less extended than cold gas, and the tidal features are not as strong (The disc scale length $R_d$ for cold gas is 3 times the $R_d$ for stars). As young stars are a small portion of the total stellar population we expect that using data from all of the stars improves the results.  Nevertheless, measuring the kinematics of stars is more challenging in faint tidal features. In real data, young stars are traced by HII regions radiating H$\alpha$, a relatively easy-to-measure emission line. With optical and near infrared imaging we can measure the morphology of the total stellar population. We plan to do a few tests on the total stellar population in morphology and young stars in kinematics, to examine possible improvements in the results (see \S \ref{sec:outlook}).

\subsection{Poor-Convergence, Retrograde and Polar Tests}

\begin{figure*}
\subfloat[]{\label{fig:retro_retro_GADGET_gas}\includegraphics[width=0.25\textwidth]{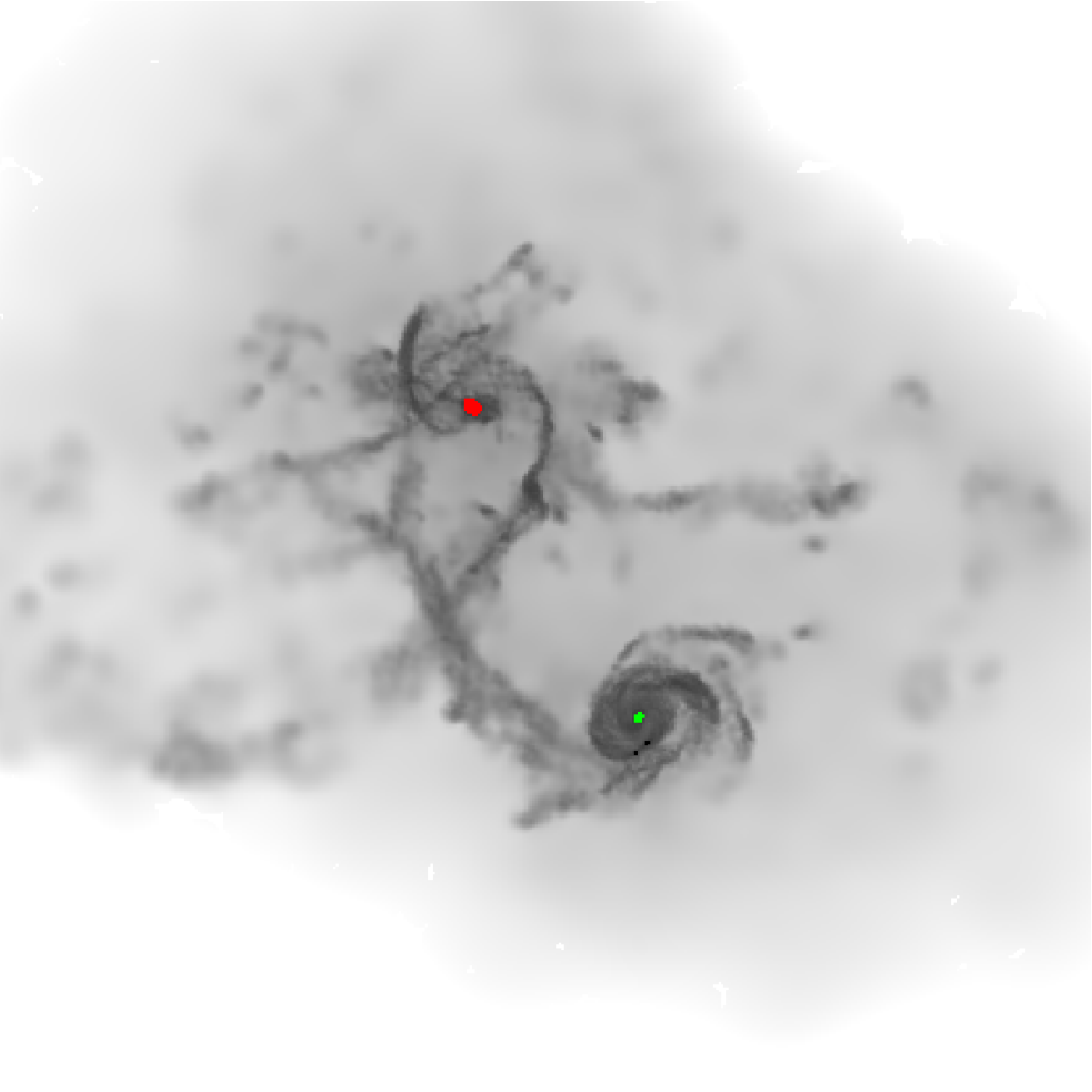}}
\subfloat[]{\label{fig:retro_retro_GADGET_newstars}\includegraphics[width=0.25\textwidth]{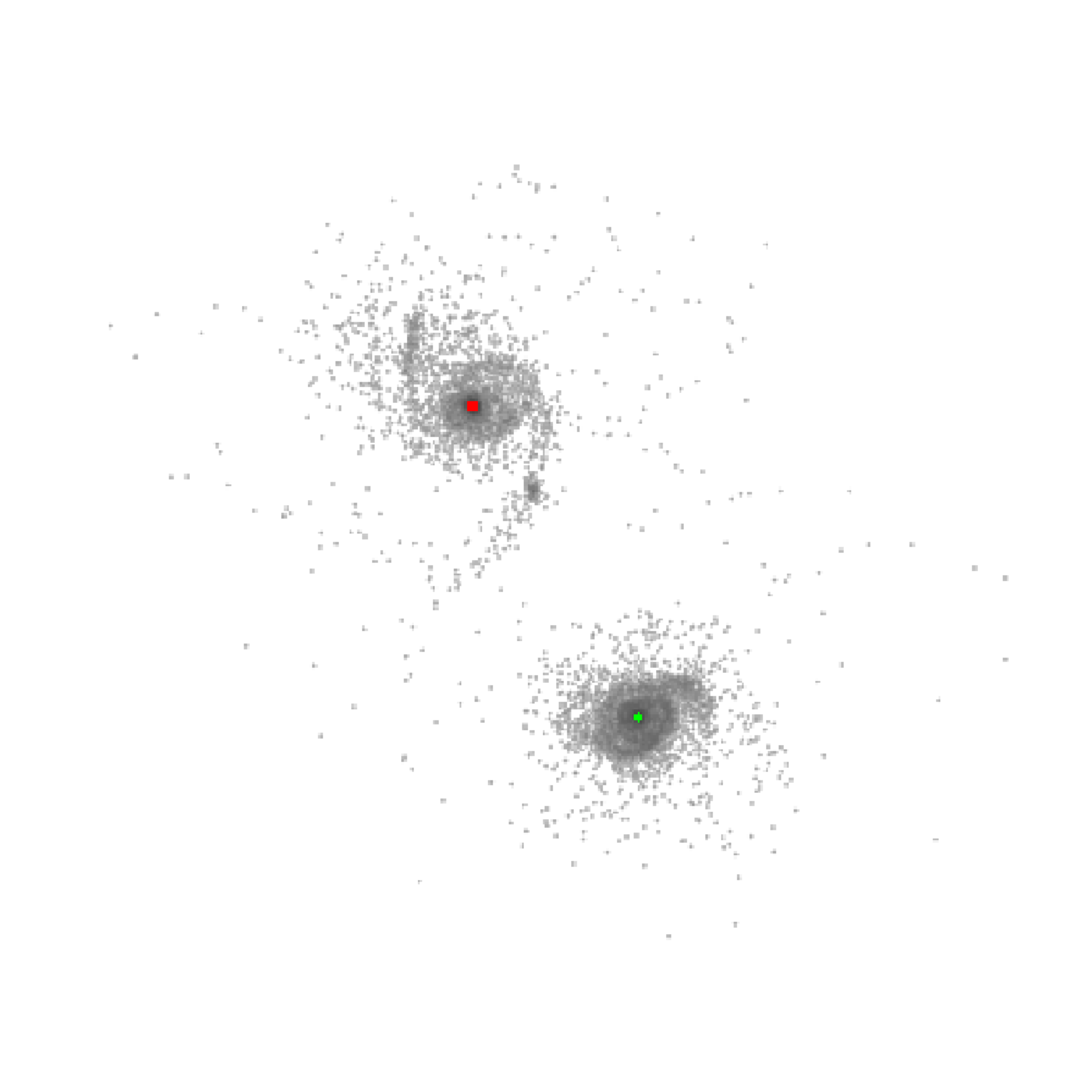}}
\subfloat[]{\label{fig:retro_retro_GADGET_oldstars}\includegraphics[width=0.25\textwidth]{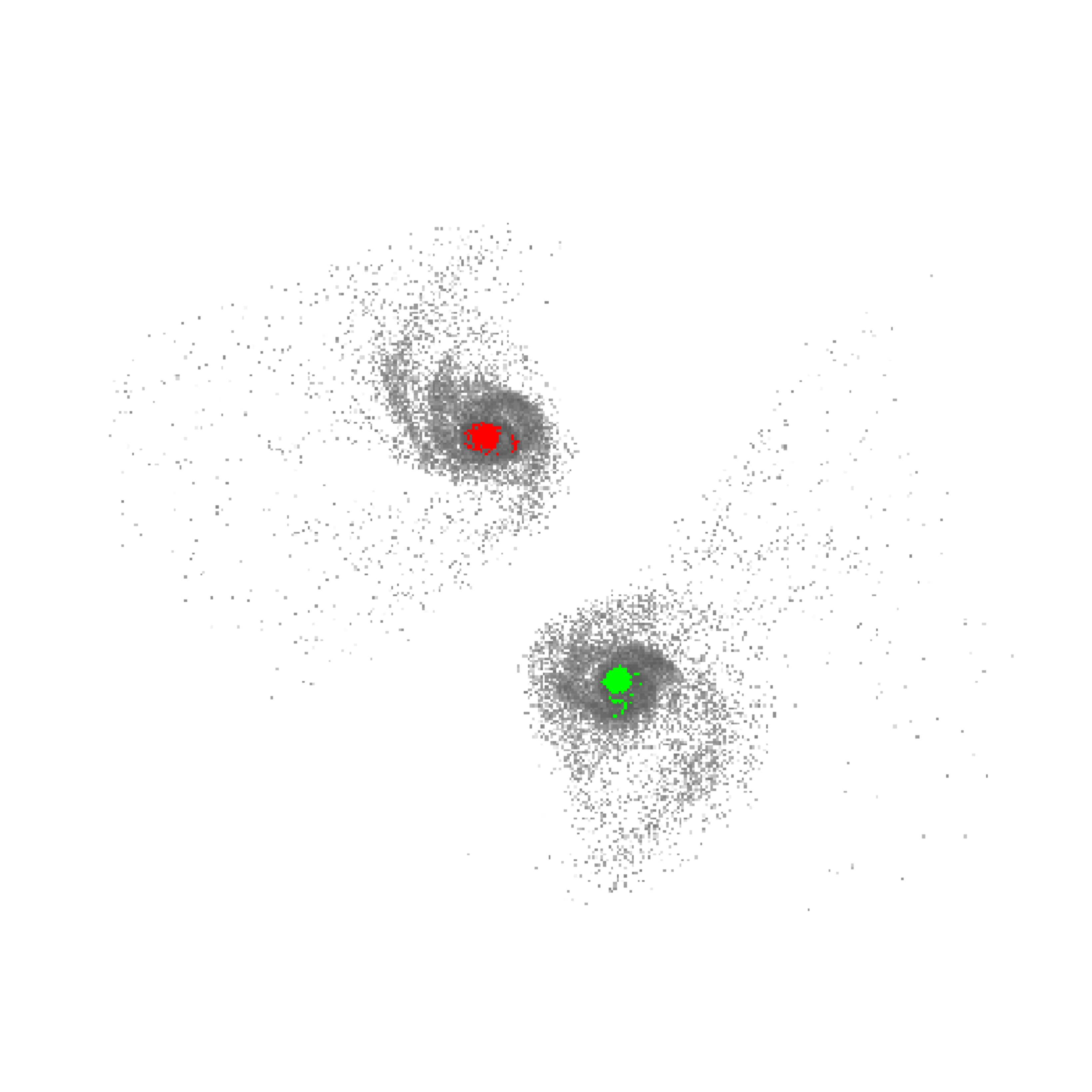}}
\subfloat[]{\label{fig:retro_retro_Identikit_correct}\includegraphics[width=0.25\textwidth]{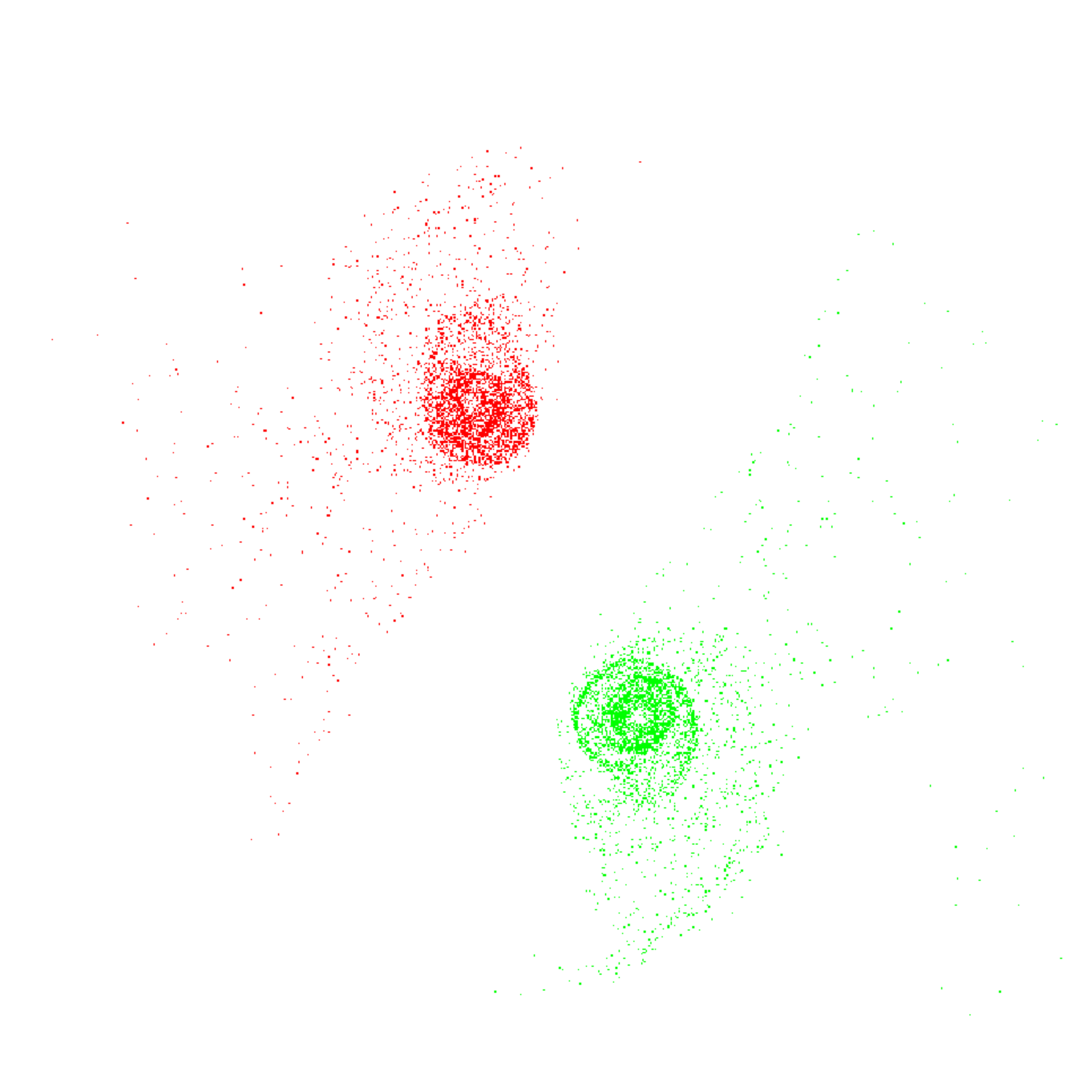}}
\caption{(a), (b), and (c) are cold gas, young stars, and old stars in GADGET simulations of the retrograde-retrograde 
test ( tests 12 and 13). Gas dissipation reshapes the gaseous tidal
  tails in a way that Identikit cannot reproduce. Young stellar population have less surface brightness
   in the outer parts of the tidal features than old stellar population. (d) shows the Identikit model with
    the same initial parameters. The tidal features in the correct Identikit model
     is a better visual match to the old stellar population (c), which is the majority of total stellar population.}
\label{fig:retro_retro}
\end{figure*}

In poor converging tests, Identikit results in random orientation of discs even in the models that are close to the best-fit model. So, when we plot the distribution of orientations in these models it looks similar to a uniform distribution (Figure \ref{fig:angledist12}). Uniform distribution of orientations leads to a cosine distribution in $\hat{n_i}.\hat{J}$. In order to quantify convergence we perform a KS test on the distribution of $\hat{n_i}.\hat{J}$ against a cosine distribution. Small KS test value means that the distribution is similar to a cosine function and there is no convergence. On the other hand, from a large KS test value we infer that the best models converged to a small range of possible orientations. A peculiar non-converging distribution (e.g. a distribution that has peaks near -1 and +1) can also give large KS test value against a cosine function, but we checked the results visually to make sure that there is no significant contamination from this type of non-converging distributions. In this work the KS test values of 0.75 and 0.30 are used as dividing values between good ($>0.75$), fair($>0.30$, $<0.75$), and poor($<0.30$) convergence.

The random orientation of discs can be the result of two scenarios. In some cases, when Identikit does not find a good combination of orientation of discs and viewing angle, it gives score of -99.00 and arbitrary orientations for discs. Alternatively, one may place the boxes in certain regions, so that many disc orientations and viewing angles populate them equally well and result in good scores. Resultingly, the distribution of orientations will look similar to a uniform distribution. For retrograde and polar tests in this work it is likely that the second scenario is causing the non-convergence.

In most of the retrograde systems the tidal features are less pronounced and closer to the center of the system. When the system has weak tidal features our automated box selection routine selects boxes close to the centers of the galaxies. The indicated phase space regions (boxes) can be populated with many configurations because of being close to the center of the galaxies. Models with poor  match to data will obtain average scores that are as good as that of models with good match. So, the best-fit model and the models close in score to the best-fit model will have a mix of parameters from well and poorly matched models, and this will make the distribution of orientations similar to that of a non-converging test. 

For one of retrograde tests, we changed the size of the boxes and tried to place them manually in places that are most similar to the choices in \citealt{Barnes:2011kb}. This attempt also failed. However, if we visually compare models with similar scores, we can distinguish a good match from a poor one. This means that Identikit algorithm does not capture all of the information one can possibly extract from visual matching. 

Retrograde mergers have other difficulties that make their modeling more complicated. There is a significant difference in the shape and position of the tidal features when we study different components of the retrograde GADGET simulation. The gas is dissipative, and when gaseous arms collide, their position, velocity, and shape starts to offset from those of the collisionless stars. Because Identikit does not have a gaseous component, it does not reproduce these differences (Figure \ref{fig:retro_retro}). Test particles better represent the collisionless stars. However, we are only testing young stars, which are a small fraction of all stars in these GADGET simulations. They are not extended as much as the total stellar population, and they have less similarity to Identikit models. One solution may be to match Identikit models to the total stellar population (Figure \ref{fig:retro_retro}). However, measuring the stellar kinematics requires high signal to noise spectroscopy which is hard to obtain in faint tidal features. We plan to test total stellar population in morphology and young stars in kinematics simultaneously to see if we can obtain better results. (See \S \ref{sec:outlook})

We think that changing the number of massive or test particles in Identikit simulations will not improve the convergence to a great extent. The features we try to match are large scale tails and bridges which are not significantly affected by the resolution of the simulations. Besides, more test particles improves the scores of all models, including models with poor match. The higher score would be due to more test particles populating the user selected phase space boxes. The scores for all models will become higher; however, what matters is the difference in scores, and higher scores in the whole score map will not improve a poor convergence.

\subsection{Outlook}
\label{sec:outlook}

We can use boxes based on morphology only to improve the constraints where morphology is more complete that the kinematics data. Tidal features are often faint. Usually they host star forming regions, though we may find no HII regions in particular areas. In case of no H$\alpha$ emission, the low continuum level of stellar light makes it very expensive to obtain uniform velocity information from stars in the outskirts of merging galaxies. We have a similar problem in our mocked data sets where the young stellar populations are less extended than the total stellar population, so they have weaker tidal features. In this work we assumed to have the H$\alpha$ kinematics from the young stellar groups, so we used young stars for matching both morphology and kinematics. Nevertheless, there is more information in the morphology of the total stellar population, and imaging of all stars is not as difficult. In Identikit we can put phase space boxes that only occupy a morphological region, and puts no constraints on velocity. In future, we can use these boxes in regions where tidal tails are present in the image but no velocity data is available. Using this type of boxes with images of all stars would have improved the constraints in young stars tests in this work.

Identikit can be improved by adding a capability that down-weights test particles which fall in regions with no features. While modeling a pair of interacting galaxies, an experienced user can select regions in phase space that no model particles should be present in a good match. Currently, there is no penalty for having particles where they are not supposed to be, and this causes some of the models with poor match to obtain good scores. We suggest to add a new type of box called ``penalty box'' to Identikit. The user puts penalty boxes in phase space regions where no particles should be present. If test-particles lay on these regions they decrease the score of the corresponding model, which then may eventually exclude some of the poor matches, and partly break the degeneracy.

Moreover, it seems that tidal features are not sufficient to constrain the orientation of discs in retrograde and polar mergers, so it is likely that using the information in the kinematics of the cores of the interacting galaxies would better constrain the initial orientation of discs. Often, an expert in modeling mergers can exclude many of the possible disc orientations by looking at the rotation curves and morphology of the discs. In the current version of Identikit the scores are calculated for a complete sphere of possible disc orientations. We suggest to improve Identikit by making the user capable of restricting the range of disc orientations to be searched. This will improve search speed, and will make it more likely to find a good match as the best-fit model.

\subsection{Summary}

In this work we have developed an automated method for modeling pairs of interacting galaxies using Identikit. Our method measures the initial conditions of major galaxy mergers and provides the uncertainty for each measurement. We tested this method using an independent set of hydrodynamical simulations of encountering galaxies. We performed the tests on galaxy merger models with various encounter parameters in eccentricity, pericentric separation, time since pericenter, viewing angle, and initial orientation of discs. We also tested both cold gas and young stars in order to check the result of using different velocity tracers for kinematics data. We found that:

\begin{enumerate}

\item We can group test results into good, fair, and poor convergence, based on the distribution of initial disc orientations in models within 1 $\sigma$ of the best fit model.
\item For all of the good and fair converged tests we recover the time since pericenter, eccentricity, pericentric distance, viewing angle, and initial disc orientations, within 30\% of the explored range from the correct value. For some parameters there are systematic offsets that can be corrected for measurements of real data
\item The tests performed on the edge-on systems result in less biased initial conditions compared to the tests on face-on systems.
\item We do not find acceptable convergence for any of the retrograde or polar tests. This may be due to the fact that we only use the tidal features to find the best match, and tidal features are not as strong in retrograde and polar systems or when we only study the young stars. The using of rotation curves as additional constraint on the system, and looking at the total stellar populations may improve these results. Besides, stripping or dissipation of cold gas may limit our tidal-tail finder.

Based on these results, our automated method can be used to find the initial conditions of prograde major galaxy mergers observed with both HI and H$\alpha$ emission lines. However, we need to improve our modeling tool for retrograde and polar systems.

\end{enumerate}

\subsection{Acknowledgments}
This project was supported in part by the STScI DDRF. This work used the Extreme Science and Engineering Discovery Environment (XSEDE), which is supported by National Science Foundation grant number ACI-1053575 (See \citealt{Towns:2014gd}). Our special thanks go to T. J. Cox who kindly provided us with the GADGET simulations. We would like to thank Prof. Colin Norman, Dr. Ron Allen, Michael Peth, and Mohammadtaher Safarzadeh for helpful comments and discussions.

\bibliographystyle{mn2e}

\bibliography{mybib}

\label{lastpage}

\end{document}